\documentclass[sigconf]{acmart}
\usepackage{times}
\usepackage{graphicx}
\usepackage{enumitem}
\usepackage{algcompatible}
\usepackage[ruled,linesnumbered]{algorithm2e}
\usepackage{cleveref}
\usepackage{subcaption}

\usepackage[skip=5pt]{caption}
\usepackage[colorinlistoftodos,prependcaption]{todonotes}

\newcommand{\framework}{{\scshape{msRep}}\xspace}

\newcommand\blue[1]{\textcolor{blue}{\textbf{#1}}}

\begin{document}

\setcopyright{acmcopyright}
\copyrightyear{2018}
\acmYear{2018}
\acmDOI{10.1145/1122445.1122456}


\title{\framework: A Fast yet Light Sparse Matrix Framework for Multi-GPU Systems}

\author{Jieyang Chen}
\email{chenj3@ornl.gov}
\affiliation{ 
    \institution{Oak Ridge National Laboratory}
}

\author{Chenhao Xie}
\email{chenhao.xie@pnnl.gov}
\affiliation{ 
    \institution{Pacific Northwest National Laboratory}
}

\author{Jesun S Firoz}
\email{jesun.firoz@pnnl.gov}
\affiliation{ 
    \institution{Pacific Northwest National Laboratory}
}

\author{Jiajia Li}
\email{jiajia.li@ncsu.edu}
\affiliation{ 
    \institution{North Carolina State University}
}

\author{Shuaiwen Leon Song}
\email{shuaiwen.song@sydney.edu.au}
\affiliation{ 
    \institution{University of Sydney}
}

\author{Kevin Barker}
\email{kevin.barker@pnnl.gov}
\affiliation{ 
    \institution{Pacific Northwest National Laboratory}
}

\author{Mark Raugas}
\email{mark.raugas@pnnl.gov}
\affiliation{ 
    \institution{Pacific Northwest National Laboratory}
}

\author{Ang Li}
\email{ang.li@pnnl.gov}
\affiliation{ 
    \institution{Pacific Northwest National Laboratory}
}

%

\begin{abstract}
Sparse linear algebra kernels play a critical role in numerous applications, covering from exascale scientific simulation to large-scale data analytics. 
Offloading linear algebra kernels on one GPU will no longer be viable in these applications, simply because the rapidly growing data volume may 
exceed the memory capacity and computing power of a single GPU. 
Multi-GPU systems nowadays being ubiquitous in supercomputers and data-centers
present great potentials in scaling up large sparse linear algebra kernels. 
In this work, we design a novel sparse matrix representation framework for multi-GPU systems called \framework, to scale sparse linear algebra operations based on our augmented sparse matrix formats in a balanced pattern.
Different from dense operations, sparsity significantly intensifies the difficulty of distributing the computation workload among multiple GPUs in a balanced manner. 
We enhance three mainstream sparse data formats -- CSR, CSC, and COO, to enable fine-grained data distribution. 
We take sparse matrix-vector multiplication (SpMV) as an example to demonstrate the efficiency of our \framework framework. 
 In addition, \framework can be easily extended to support other sparse linear algebra kernels based on the three fundamental formats (i.e., CSR, CSC and COO).

\end{abstract}

\maketitle

\section{Introduction}

Graphics processing units (GPUs) have become the mainstream and powerful accelerators in the past decade and supported a wide spectrum of applications, such as 
applications based on direct and iterative solvers,  machine learning algorithms, etc. 
Due to advances in interconnect technology, multiple-GPU systems nowadays have been widely adopted in from the world's fastest supercomputer Summit~\cite{summit} to NVIDIA's Super-AI DGX systems~\cite{dgx}, and even desktop workstations. 
With a condensed configuration, up to 16 GPUs can be installed in a single compute node with fast interconnect such as NVLink or NVSwitch connecting them. These high-speed interconnects offer much superior  data-exchanging rate among GPUs than the conventional PCI-e solution, where data always has to be routed by CPUs. This fundamental change poses a unique opportunity for designing scalable multi-GPU algorithm for large-scale data processing.


Nevertheless, except for deep-learning tasks, few research has been conducted on optimizing performance for a multi-GPU system leveraging these new interconnect. Some work targeted  multi-GPUs~\cite{pan2017multi,ben2017groute,kreutzer2012sparse,Yang2011,abdelfattah2015high}, but due to restricted bandwidth of the PCI-e, they tend to avoid communication whenever possible, undermining the capability of the fast interconnect; a general framework that can fill this gap can be highly beneficial to the community.

To facilitate multi-GPU programming and data sharing among the GPUs, the vendors like NVIDIA has introduced Unified Memory \cite{UnifiedMemory} and NVSHMEM technology~\cite{nvnvshmem}. 
Nonetheless, the designing of malleable multi-GPU data structures while achieving workload balance have been left over at the discretion of the programmers.
In this work, we target the domain of sparse matrix operations. In particular, we introduce novel data structures for storing sparse data on multi-GPU systems. In addition, we propose a workload balancing technique for multi-GPU sparse kernels. 
Unlike dense linear algebra kernels, the sparsity feature makes it challenging to achieve balanced workload distribution.

To this end, we consider sparse matrix-vector multiplication (SpMV) kernel. SpMV is one of the most extensively utilized sparse matrix operations in big data analytics and scientific computations. 
This kernel has been widely studied in many research work~\cite{im2004sparsity,liu2015csr5,bell2009implementing,choi2010model,anzt2014implementing,ashari2014fast,kreutzer2012sparse,monakov2010automatically,yan2014yaspmv,li2013smat,greathouse2014efficient,merrill2016merge,xie2018cvr,steinberger2017globally,tan2018design,hong2018efficient,zhao2018bridging,hou2017auto,vuduc2003automatic,williams2007optimization,schubert2011hybrid} in the context of data structure design, performance optimization, compiler implementation, and hardware architecture support, as well as on shared memory, distributed systems, and GPUs.
However, to the best of our knowledge, 
few studies have been carried out in designing SpMV kernel for multi-GPU systems to take the advantage of fast GPU-GPU interconnect. Existing works on SpMV either involve CPUs to orchestrate data transfers or split the whole workload into small, independent tasks to distribute on multiple GPUs without giving careful consideration to the workload imbalance~\cite{abdelfattah2015high,guo2016performance}.
Moreover, applications with large linear systems 
make offloading an SpMV kernel on a single GPU infeasible due to its limited memory capacity.
While possible solutions of deploying out-of-core or distributed SpMV exist, these solutions suffer from slow CPU-GPU transfer rate or 
network latency. In these cases, it is challenging to guarantee high performance, especially for memory-bound SpMV kernel.
Hence, we take SpMV as the driving kernel to showcase and illustrate our sparse matrix representation framework's efficiency on multi-GPU systems.


In this work, we propose a general framework, namely \framework, consisting of augmented sparse matrix formats and balanced distribution. We demonstrate that, with our framework, sparse linear algebra kernels, in particular, SpMV, can achieve scalable performance
on multi-GPU systems. 
In addition, 
 our framework can support the existing SpMV kernels based on these formats. 
Specifically, our contributions are as follows:
\begin{itemize} [leftmargin=*]
    \item We 
    extended three popular sparse data formats, Compressed Sparse Row (CSR), Compressed Sparse Colunmn (CSC), and coordinate (COO) by storing a part of a sparse matrix with arbitrary start and end positions. The new formats, pCSR, pCSC, and pCOO, requires small additional memory to maintain the metadata and can be converted from the well-known formats swiftly.
    \item To demonstrate our framework's efficiency on multi-GPU systems, we develop an SpMV kernel for multi-GPU system called \texttt{mSPMV}.
    To achieve better scalability, \texttt{mSPMV} leverages our sparse matrix representations on multi-GPUs for efficient workload distribution. 
    \item We evaluate our \framework on two dense GPU systems: the Summit supercomputer at Oak Ridge National Laboratory and a NVIDIA V100-DGX-1 system. Experiments using matrices from the \emph{SuitSparse Matrix Collection}~\cite{sparse-matrix-collection} shows that \texttt{mSPMV} can achieve 5.5X speedup using six GPUs on Summit and 6.2X speedup using eight GPUs on the NVIDIA V100-DGX-1 system.
\end{itemize}

\vspace{4pt}\noindent The rest of this paper is organized as follows. In Section~\ref{backgound} we provide background of this work with a discussion about three popular sparse data formats and use SpMV as an example to illustrate workload imbalance due to data distribution. In Section~\ref{design}, we discuss in detail the design of our sparse matrix representation framework together with the design of the multi-GPU SpMV. We propose several implementation optimizations for SpMV on multi-GPU systems in Section~\ref{opt}. We report our experimental results in Section~\ref{experiements}. We summarize our observations in Section~\ref{discussion}. We give an overview of related works in Section~\ref{related} and draw our conclusion in Section~\ref{conclusion}.


\begin{table}[h]
\centering
\caption{Notation in Algorithms and Formulations.}
\label{notation}
\begin{tabular}{|c|l|}
\hline
 $m$         & Number of rows in the input matrix. \\ \hline
 $n$         & Number of columns in the input matrix. \\ \hline
 $nnz$   & Number of non-zero elements in the input matrix. \\ \hline
 $np$   & Total number of partitions. \\ \hline
 \end{tabular}
\end{table}

\begin{figure}[h]
    \vspace{-2mm}
    \centering
    \includegraphics[width=0.12\textwidth]{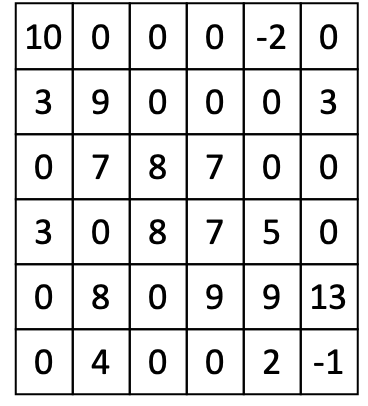}
    \caption{A sparse matrix where a large portion are zero elements.}
    \vspace{-2mm}
    \label{example-mat}
\end{figure}

\section{Background}
\label{backgound}

We use the example matrix in \textbf{Fig. \ref{example-mat}} to illustrate different storage formats and their SpMV algorithms. We also analyze the imbalance issue on multi-GPU systems for sparse matrix operations. \textbf{Table \ref{notation}} shows the related notations for this paper, for a $m \times n$ matrix with $nnz$ non-zero elements.

\begin{figure}[h]
    \centering
    \includegraphics[width=0.4\textwidth]{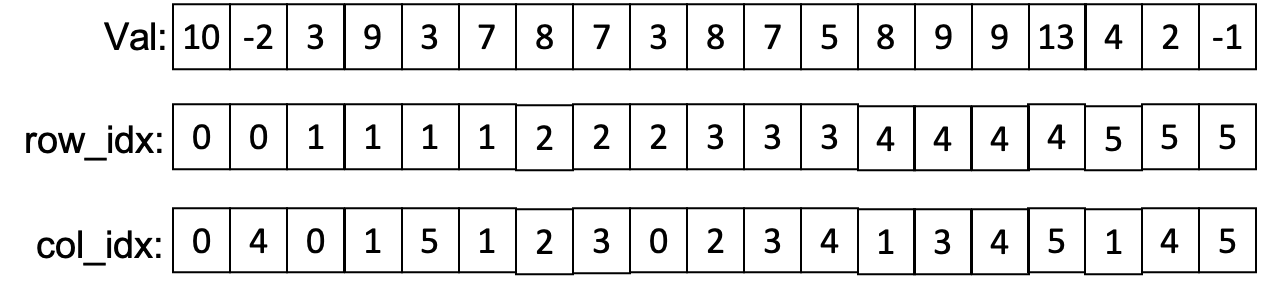}
    \caption{Coordinate (COO) sparse matrix format.}
    \label{coo}
\end{figure}

\begin{figure}[h]
    \centering
    \includegraphics[width=0.4\textwidth]{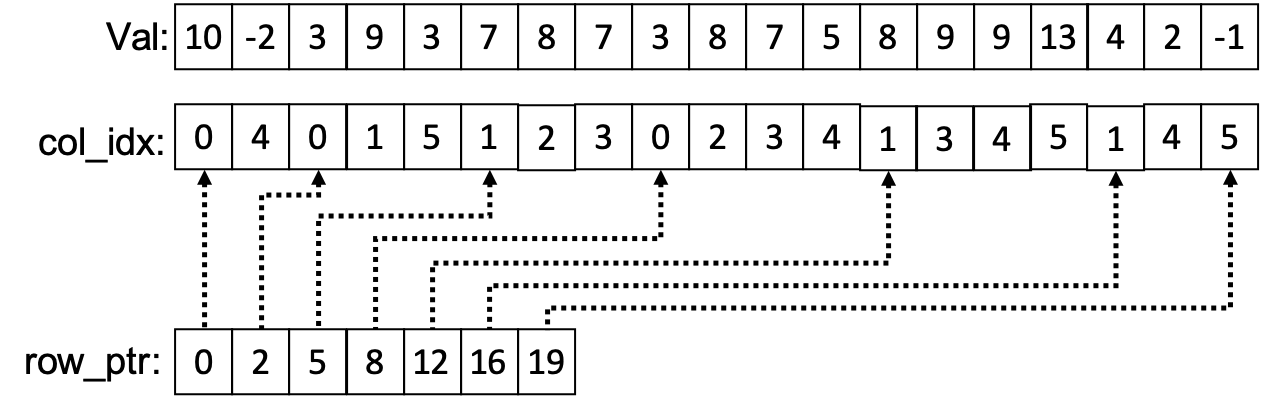}
    \caption{Compressed Sparse Row (CSR) sparse matrix format.}
    \label{csr}
\end{figure}

\begin{figure}[h]
    \centering
    \includegraphics[width=0.4\textwidth]{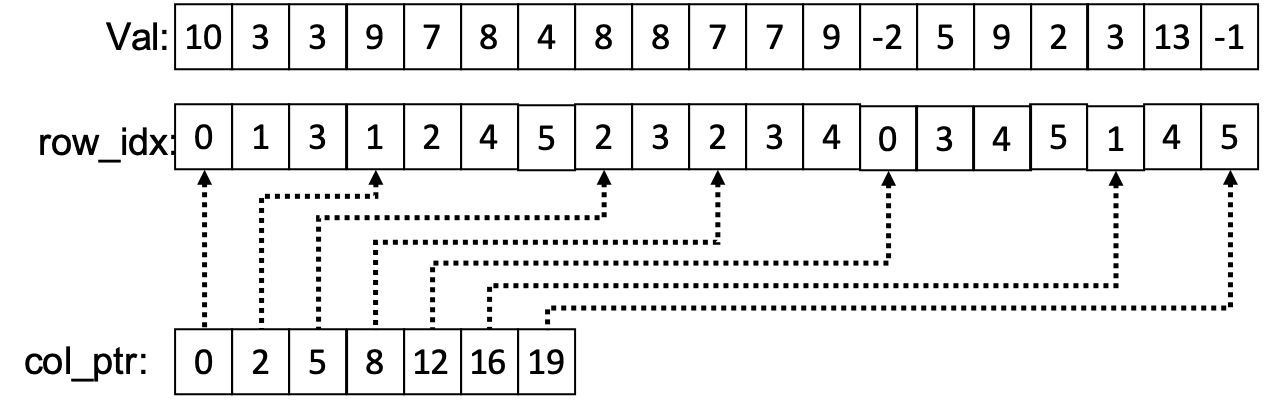}
    \caption{Compressed Sparse Column (CSC) sparse matrix format.}
    \label{csc}
\end{figure}

\subsection{Mainstream Sparse Matrix Storage Formats}

\subsubsection{COO Format}
Coordinate (COO) format is the most straightforward sparse matrix format by storing only non-zero elements along with their indices and values.
\textbf{Fig. \ref{coo}} shows the data structure of the COO format to store the example sparse matrix (\textbf{Fig. \ref{example-mat}}). 
Three $nnz$-sized arrays are used. \texttt{val} stores the values of non-zero elements. \texttt{row\_idx} and \texttt{col\_idx} store the row and column indices corresponding to each non-zero element. 

\subsubsection{CSR Format}
Compressed Sparse Row (CSR) format is a more compressed format compared to the COO format.  
\textbf{Fig. \ref{csr}} gives the data structure of the CSR format using the same example sparse matrix (\textbf{Fig. \ref{example-mat}}). 
The same with COO, \texttt{val} stores the non-zero values and \texttt{col\_idx} stores their column indies. \texttt{row\_ptr} compresses row indies by pointing the row starting position.

\subsubsection{CSC Format}
Compressed Sparse Column (CSC) is similar to the CSR format, which is also compressed but on columns.
\textbf{Fig. \ref{csc}} shows the CSC format where \texttt{val} stores the non-zero values,  \texttt{row\_idx} stores row indices, and \texttt{col\_ptr} stores the pointers of starting position of column indices.
The CSC format of a matrix $A$ is the same with the CSR format of its transposed matrix $A^T$.

\subsection{Sparse Matrix-Vector Multiplication}
Sparse Matrix-Vector Multiplication (SpMV) operation is the most popular operation of sparse linear algebra and has broad applications. In this work, we take SpMV as an illustration example to show the benefit of our framework \framework.

\textbf{Algorithm \ref{spmv-simple}} calculates the SpMV using CSR format.
This algorithm loops all rows of a sparse matrix; then for each row, loops all the non-zeros inside.
According to the $col\_idx[j]$ index to locate the $x$ value to do the product with non-zero value $val[j]$. 
All the partial products on row $i$ will be summed and used to update $y[i]$ for the final output.
SpMV algorithm based on CSC format is to switch the role of $x$ and $y$.
SpMV algorithm using COO format is not hard to image: only one loop for all the $nnz$ non-zero elements while its corresponding column index to locate the corresponding $x$ element then do the product and update its counterpart output $y$ element by indexing with its row index.

\SetKwInOut{KwInOut}{In/Out}
\SetKwInOut{KwIn}{In}
\SetKwInOut{KwOut}{Out}
\begin{algorithm}[]
\caption{CSR-based SpMV: $y = \alpha Ax + \beta y$}
\label{spmv-simple}
\KwIn{spares matrix A ($m \times n$): val, col\_idx, and row\_ptr}
\KwIn{dense vector x ($n$)}
\KwIn{scalar $\alpha$ and $\beta$}
\KwInOut{dense vector y ($m$)}
\For{$i=1$ to $m$}{
	\For{$j=row\_ptr[i]$ to $row\_ptr[i+1]$}{
		$y[i] = \alpha * val[j] * x [col\_idx[j]] + \beta * y[i]$
	}
}
\end{algorithm}

\subsection{Imbalance issue on multi-GPU systems}
To motivate our work we show how workload distribution strategies can impact the performance of a sparse matrix operation.
Take SpMV as the example, it is generally a memory bound computation on mainstream architectures as its flops-to-bytes ratio is roughly $O(1)$. 
Thus, the cost of loading input data is a main dominant factor of its performance due to its relatively large memory footprint compared to the two vectors and the potential data reuse due to cache and memory hierarchy~\cite{chen2019tsm2}.
For the input matrix, each element is only used once during the entire computation, so the main cost of SpMV comes from accessing the nonzero elements.
For other sparse matrix operations, sparse matrix-sparse vector and sparse matrix times multiple dense vectors have similar behavior with SpMV. Even for sparse matrix-dense/sparse matrix operation, due to the potential data reuse of the right matrix, the memory access of sparse matrix still playing an important role.

\begin{figure}[]
    \centering
    \includegraphics[width=0.35\textwidth]{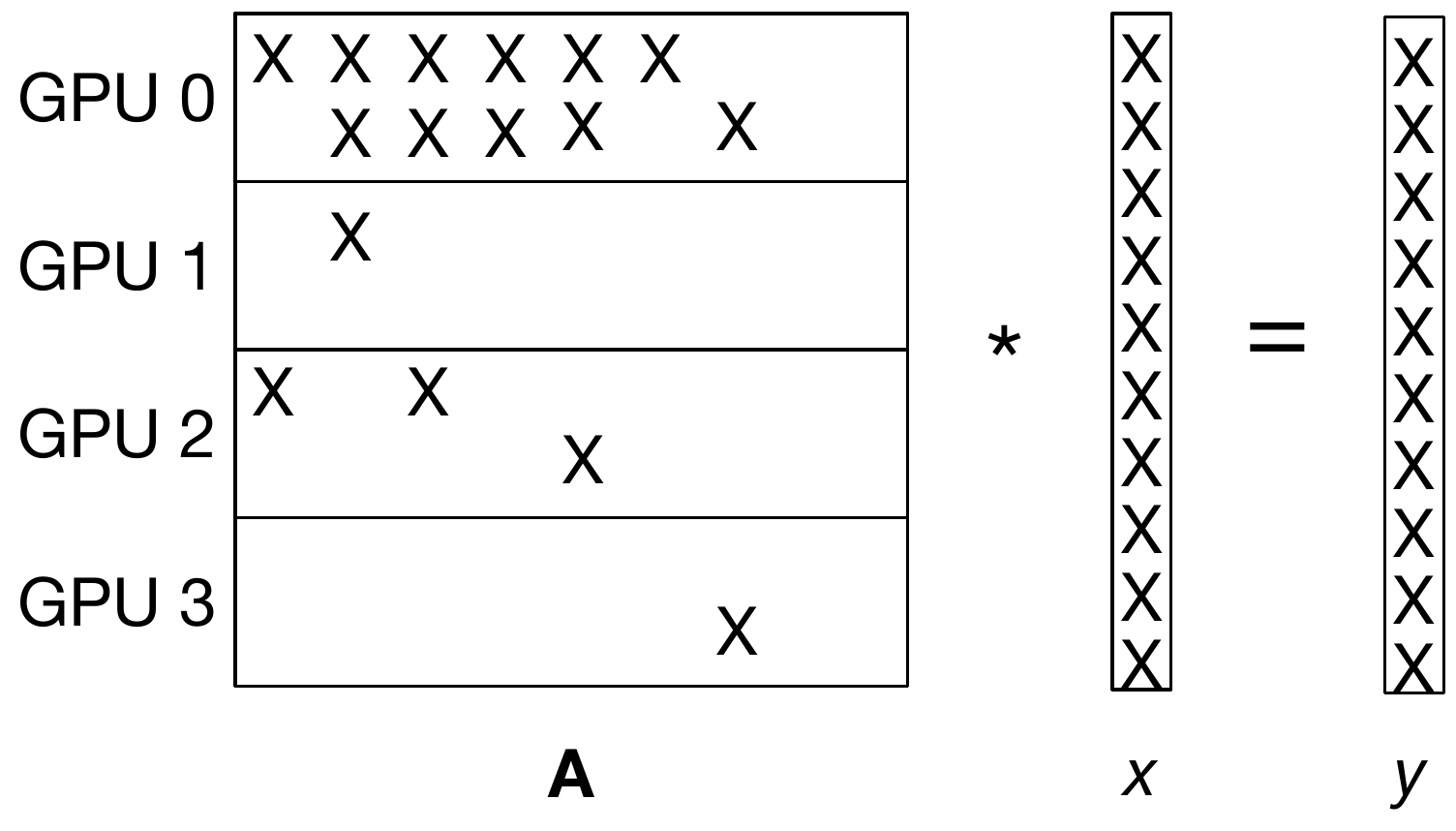}
    \caption{Naive workload distribution of SpMV with non-uniformly distributed input matrix elements. }
    \label{baseline-fig}
\end{figure}

\begin{figure}[]
    \centering
    \includegraphics[width=0.4\textwidth]{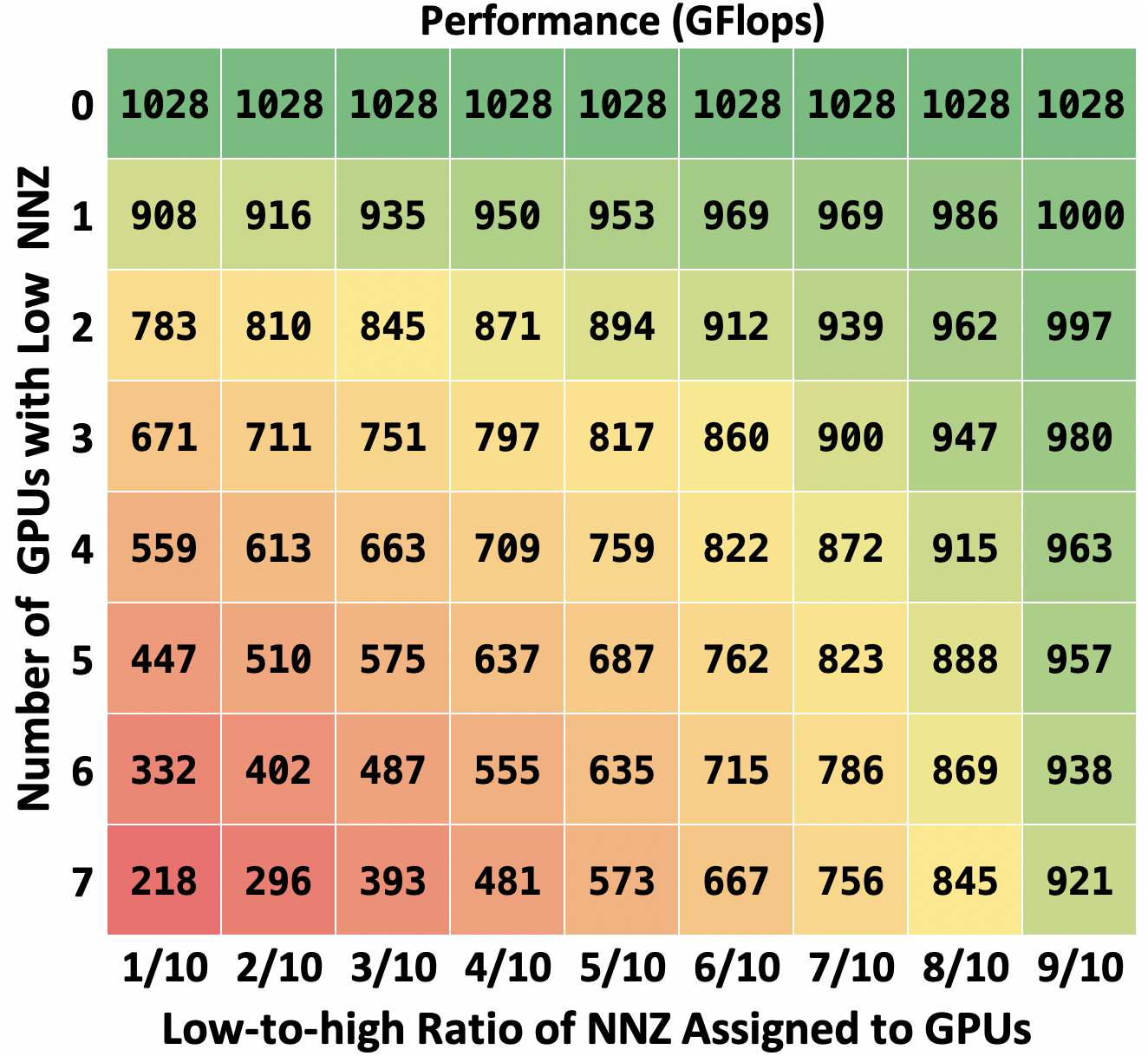}
    \caption{Imbalanced non-zero elements distribution among GPUs can cause performance degradation in SpMV. For example, if 4 of the total 8 GPUs are assigned with only 1/10 of the none-zeros assigned to the other 4 GPUs, the overall performance would reduce to about half (559/1028).
    }
    \label{baseline}
\end{figure}

For the dense matrix multiplication, a commonly used strategy to distribute elements in the input matrix is simply dividing the matrix into row blocks and then assigning each of them to different GPUs for computation.
Distributing rows evenly among GPUs would leads to good workload balancing and performance.
However, when applied to sparse matrix, this kind of workload distribution will not work well without considering the sparsity of the matrix.
Since calculations on zeros are unnecessary, they are usually omitted, which leads to workload imbalance as the number of zeros may varies in between row blocks as shown in \textbf{Fig. \ref{baseline-fig}}.
For SpMV, the workload of each GPU is proportional to the number of non-zero elements ($nnz$) rather than the number of rows ($m$). 
Workload imbalance in SpMV could greatly impact the overall SpMV performance. \textbf{Fig. \ref{baseline}} shows the benchmarking results of using the straightforward distributed strategy in SpMV. 
We generate input matrices with different kinds of non-zeros element distribution that lead to imbalanced $nnz$ on each GPUs.
To simplify, the distribution leads to two kinds of workload among GPUs. One kind of workload has higher number of $nnz$ than the other ones. The ratio of $nnz$ between low-to-high is shown in the x-axis. The test is conducted on a NVIDIA V100-DGX-1 system.

\section{\framework Framework}
\label{design}
We introduce our \framework framework in this section with our proposed three enhanced formats, pCSR, pCSC, and pCOO and workload management. 

\subsection{Challenges}

\begin{figure}[]
    \centering
    \includegraphics[width=0.3\textwidth]{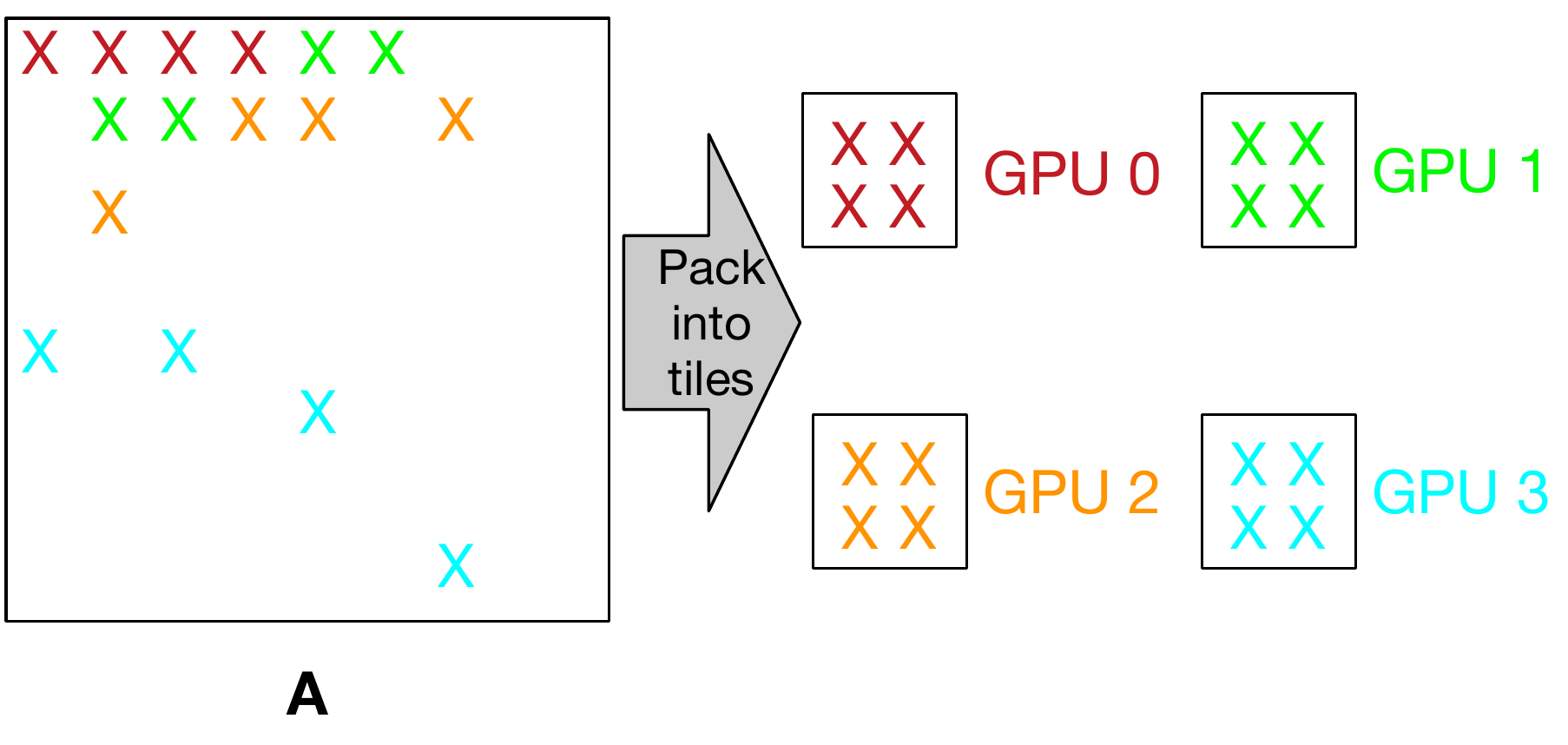}
    \caption{Ideal SpMV workload distribution based on none-zeros.}
    \label{balanced-fig}
\end{figure}

Distributing the input matrix according to $nnz$ (shown in \textbf{Fig. \ref{balanced-fig}}) is the most straightforward and efficient way to divide the entire workload in finer granularity to achieve better workload balance. 
As the focus of this work is exploiting parallelism across GPUs, we choose to leverage existing well-optimized works~\cite{bell2009implementing,hong2018efficient,steinberger2017globally,merrill2016merge,greathouse2014efficient,ashari2014fast,BASMAT,choi2010model,hou2017auto,tan2018design,monakov2010automatically,liu2015csr5,yan2014yaspmv,bell2009implementing,anzt2014implementing} to handle workload on each single GPU.
Popular state-of-the-art works and libraries support at least one of the three mainstream sparse matrix storage formats: CSR, CSC, and COO.
So, compatible with these three format can allow us not only leverage existing state-of-the-art kernels but also benefit from them in the future.
However, making the workload distribution efficient and compatible is non-trivial. 

\subsection{Fine-Grain Workload Distribution}
\label{part}
\subsubsection{pCSR}
We first propose a data format called partialCSR (pCSR).
pCSR can easily represent a subset of non-zero elements in a sparse matrix while preserving all necessary element distribution information. It can be converted from CSR format efficiently.
Once CSR is partitioned into pCSR, it can be used by CSR-based SpMV and other kernels without overhead.

\begin{figure}[h]
    \centering
    \includegraphics[width=0.4\textwidth]{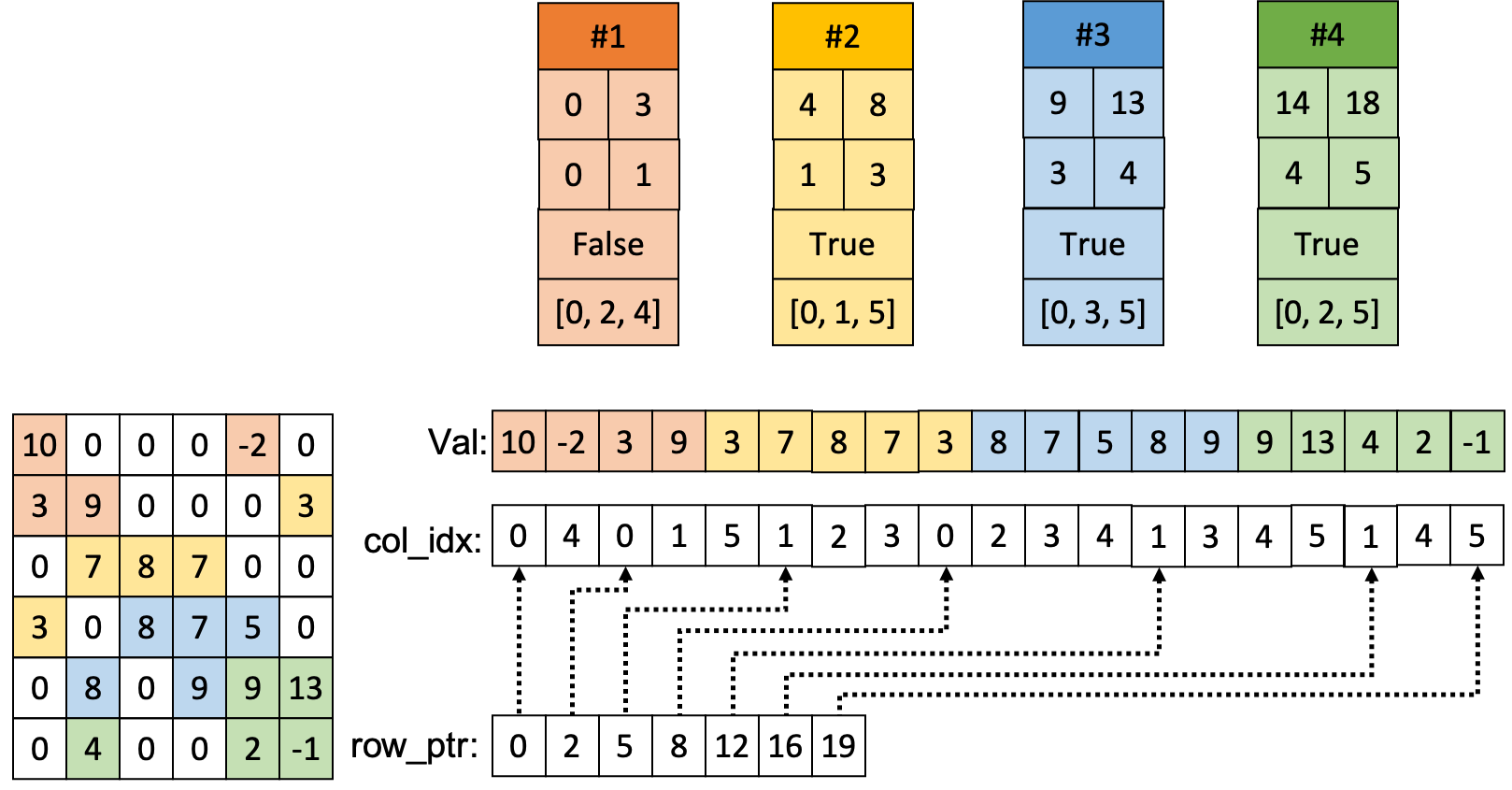}
    \caption{Example of partitioning a matrix into four parts using pCSR. }
    \label{pCSR-example}
\end{figure}

\textbf{Fig. \ref{pCSR-example}} shows the data structure of pCSR. 
To enable efficient conversion to/from CSR format, the most straightforward way is to avoid data copy. 
So, to represent a subset of non-zero elements in a sparse matrix, we use two index values (i.e., \texttt{start\_idx} and \texttt{end\_idx}) to mark the starting and ending position in the non-zero array of CSR. 
The storage cost is $O(1)$. 
However, maintaining a local row pointer array is necessary for SpMV kernels to perform correctly. 
The storage cost is proportional to the number of rows in the partition, which varies depending on non-zero elements distribution.
The total cost is no more than $O(m)$.
This local row pointer array can be calculated efficiently and will be discussed later. 
Also, we use the two index values (i.e., \texttt{start\_idx} and \texttt{end\_idx}) to mark the starting and ending position in the column index array, so no additional cost is introduced.
Since elements in the same row could be distributed into multiple pCSRs, we also maintain a flag (i.e., \texttt{start\_flag}) to mark if the first row maintained by the current pCSR is partial or not.
The last row does not need to be flagged as it can be inferred from the \texttt{start\_flag} of the next pCSR.
Finally, for merging multiple pCSR into one CSR it is necessary to maintain two indices that stores the start and end row index in the global view.

\begin{algorithm}
\caption{Converting from CSR to pCSR}
\label{csr-pCSR}
\KwIn{CSR matrix A}
\KwIn{Number of Non-zeros in A: $nnz$ }
\KwIn{Number of pCSRs: $np$}
\KwOut{pCSR matrix pA[np]}

\For{$i=1$ to $np$}{
    pA[i].start\_idx=$\lfloor i * nnz / np \rfloor$ \\
    pA[i].end\_idx=$\lfloor (i+1) * nnz / np \rfloor -1$ \\
    pA[i].start\_row=BinarySearch(A.row\_ptr, pA[i].start\_idx) \\
    pA[i].end\_row=BinarySearch(A.row\_ptr, pA[i].end\_idx) \\
    \eIf {pA[i].start\_idx > A.row\_ptr[pA[i].start\_row]}{
        pA[i].start\_flag = True \\
    }{   
        pA[i].start\_flag = False \\
    }
    \For{$j=1$ to pA[i].end\_row - pA[i].start\_row}{
        pA[i].row\_ptr[j] = A.row\_ptr[pA[i].start\_row + j] - pA[i].start\_idx \\
    }
}
\end{algorithm}

\textbf{Algorithm \ref{csr-pCSR}} shows how to convert CSR to pCSR format. 
The main cost comes from searching the start and end indices using Binary Search (i.e., $O(log(m))$) and computing the local row pointers (i.e., $O(pA[i].end\_row - pA[i].start\_row)$). 
The total cost is $O(np*log(m) + m)$. 
The former one can be efficiently done on CPUs and the local row pointers can be computed using GPUs. 
We will show time cost comparison of the partitioning with and without GPUs.
Finally, each individual partition can be generated independently so the partitioning process can be efficiently parallelized.

\begin{algorithm}
\caption{Using pCSR on CSR-based SpMV kernels}
\label{pCSR-kernel}
\KwIn{CSR matrix A}
\KwIn{dense vector x ($n \times 1$)}
\KwIn{scalar $\alpha$ and $\beta$}
\KwInOut{dense vector y ($m \times 1$)}
\KwIn{Number of pCSRs: $np$}
\KwIn{pCSR matrix pA[np]}
Allocate space to hold partial results py[np]\\
\blue{/* Run in parallel on multi-GPU */} \\
\For{$i=1$ to $np$}{
    val = A.csr[pA[i].start\_idx] \\
    row\_ptr = pA[i].row\_ptr \\
    col\_idx = A.col\_idx[pA[i].start\_idx] \\
    Launch: py[i]=<csrSpMVKernel>(val, row\_ptr, col\_idx) \\
}

\For{$i=1$ to $np$}{
    \If{pA[i].start\_flag}{
         tmp = y[pA[i].start\_row] \\
    }
    memcpy: y[pA[i].start\_row] $\leftarrow$ py[i] \\
    \If{pA[i].start\_flag}{
        y[pA[i].start\_row] -= tmp * $\beta$ \\
    }
}

\end{algorithm}

\textbf{Algorithm \ref{pCSR-kernel}} shows how to launch CSR-compatible SpMV kernel using pCSR format. 
We can see pCSR can be converted to CSR without overhead when invoking SpMV kernels. 
This ensures that all existing and future CSR-compatible SpMV kernel on single GPU can use our framework. \textbf{Line 9 - 17} shows how to correctly merge a series of partial results into a final result. 
We will further discuss how it can be efficiently done in section \ref{opt}.

\subsubsection{pCSC}
\begin{figure}[h]
    \centering
    \includegraphics[width=0.5\textwidth]{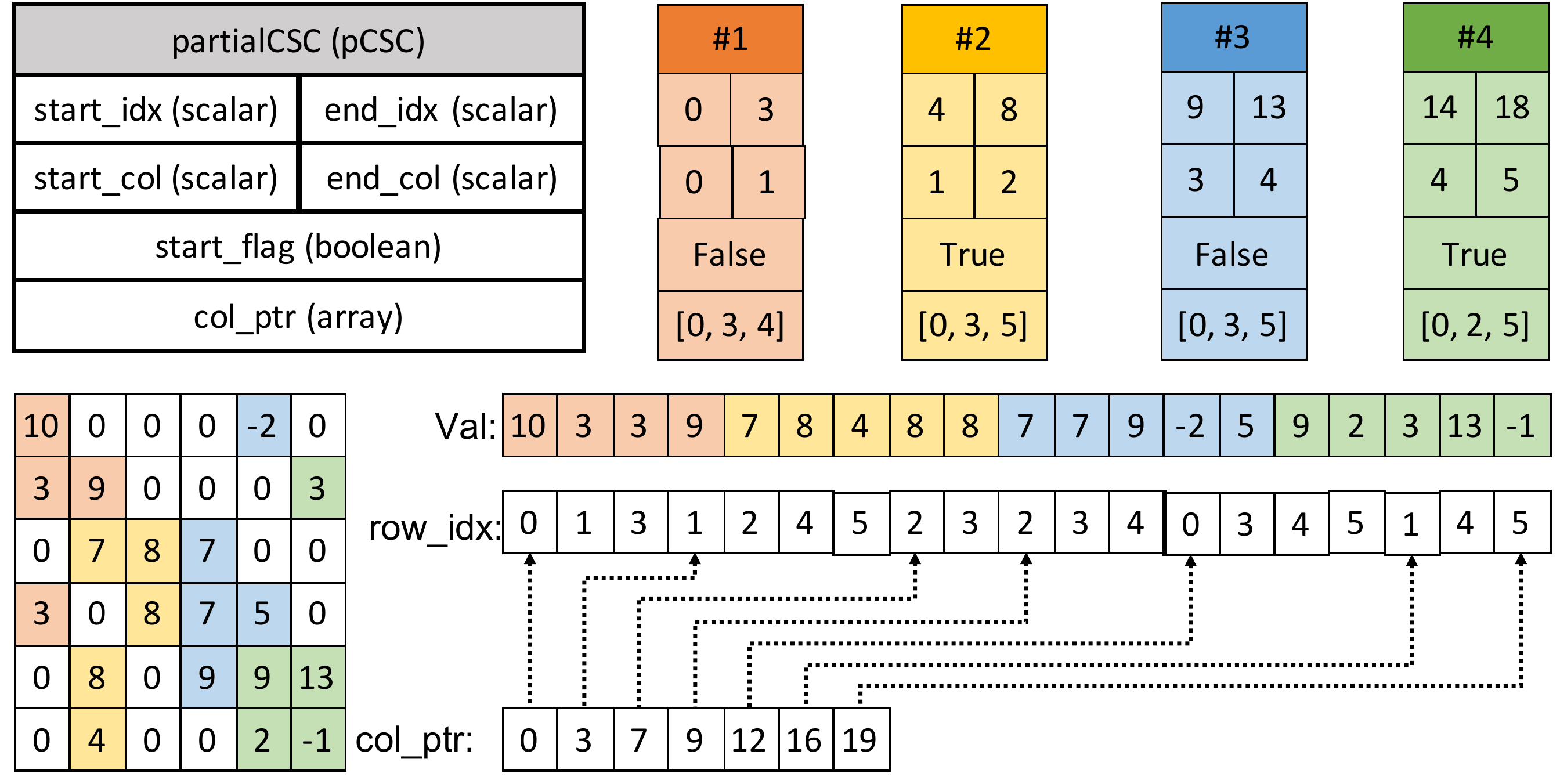}
    \caption{Example of partitioning a matrix into four parts using pCSC. 
    }
    \label{pcsc-example}
\end{figure}

Similar to pCSR, we also propose partialCSC (pCSC) for partitioning a sparse matrix that is stored in CSC format as shown in \textbf{Fig. \ref{pcsc-example}}.
\text{Algorithm \ref{csc-pcsc}} shows how to efficiently convert CSC to pCSC format. 
It is easy to see that the overall cost of the conversion is similar to pCSR format. 
The total cost is $O(np*log(n) + n)$. 
The algorithm can also be efficiently parallelized. \text{Algorithm \ref{pcsc-kernel}} shows how to use pCSC in CSC-based SpMV kernels. Again, we will discuss how to do efficient result merging in section \ref{opt}.


\begin{algorithm}
\caption{Converting CSC to pCSCs}
\label{csc-pcsc}
\KwIn{CSC matrix A}
\KwIn{Number of pCSCs: $np$}
\KwIn{pCSC matrix pA[np]}
\For{$i=1$ to $np$}{
    pA[i].start\_idx=$\lfloor i * nnz / np \rfloor$ \\
    pA[i].end\_idx=$\lfloor (i+1) * nnz / np \rfloor -1$ \\
    pA[i].start\_col=BinarySearch(A.col\_ptr, pA[i].start\_idx) \\
    pA[i].end\_col=BinarySearch(A.col\_ptr, pA[i].end\_idx) \\
    \eIf {pA[i].start\_idx > A.col\_ptr[pA[i].start\_col]}{
         pA[i].start\_flag = True \\
    }{   
        pA[i].start\_flag = False \\
    }
    \For{$j=1$ to pA[i].end\_col - pA[i].start\_col}{
        pA[i].col\_ptr[j] = A.col\_ptr[pA[i].start\_col + j] - pA[i].start\_idx \\
    }
}
\end{algorithm}

\begin{algorithm}
\caption{Launching CSC-based SpMV kernel using pCSC}
\label{pcsc-kernel}
\KwIn{CSC matrix A}
\KwIn{dense vector x ($n \times 1$)}
\KwIn{scalar $\alpha$ and $\beta$}
\KwInOut{dense vector y ($m \times 1$)}
\KwIn{Number of pCSCs: $np$}
\KwIn{pCSC matrix pA[np]}
Allocate space to hold partial results py[np]\\
\blue{/* Run in parallel on multi-GPU */} \\
\For{$i=1$ to $np$}{
    val = A.csc[pA[i].start\_idx] \\
    col\_ptr = pA[i].col\_ptr \\
    row\_idx = A.row\_idx[pA[i].start\_idx] \\
    Launch: py[i]=<cscSpMVKernel>(val, col\_ptr, row\_idx) \\
}

\For{$i=1$ to $np$}{
    sum\_y += py[i] \\
}
y = 
\end{algorithm}

\subsubsection{pCOO}
Finally, we propose partialCOO (pCOO) format for partitioning COO format-based sparse matrices.
To void the cost of element reordering, we choose to generate partitions by dividing the input into consecutive non-zero elements.
Partitioning COO is slightly different than CSR and CSC since the elements can be sorted or unsorted.
The whether or not the COO sorted will matters to how much information we would be able to know about a partition.
If the elements are sorted, for example sorted by rows, then it is possible to efficiently find the start and end row index correspond to the partition so that we can locate corresponding row of result vector it will calculate and facilitate the partial merging process.
However, if the elements are unsorted, we can only assume that the elements in a particular partition can spread among the entire matrix without knowing row or column range. 
This could bring extra memory cost for storing partial results and time for merging partial results.
For simplicity, we assume the elements are sorted by rows in this paper.
\text{Algorithm \ref{coo-pcoo}} shows how to efficiently convert COO to pCOO format. It is easy to see that the overall cost is $O(np*log(m))$ assuming the non-zeros are already sorted by the row index. If sorted by column index the cost is $O(np*log(n))$. The algorithm can be efficiently parallelized into $np$ individual tasks. \text{Algorithm \ref{pcoo-kernel}} shows how to use pCOO in COO-based SpMV kernels. Again, we will discuss how to do efficient result merging in section \ref{opt}.

\begin{algorithm}
\caption{Converting COO to pCOOs}
\label{coo-pcoo}
\KwIn{COO matrix A}
\KwIn{Number of pCOOs: $np$}
\KwIn{pCOO matrix pA[np]}
\For{$i=1$ to $np$}{
    pA[i].start\_idx=$\lfloor i * nnz / np \rfloor$ \\
    pA[i].end\_idx=$\lfloor (i+1) * nnz / np \rfloor -1$ \\
    pA[i].start\_row=BinarySearch(A.row\_ptr, pA[i].start\_idx) \\
    pA[i].end\_row=BinarySearch(A.row\_ptr, pA[i].end\_idx) \\
    \eIf {pA[i].start\_idx > A.row\_ptr[pA[i].start\_row]} {
        pA[i].start\_flag = True \\
    }{   
        pA[i].start\_flag = False \\
    }
}
\end{algorithm}

\begin{figure}[h]
    \centering
    \includegraphics[width=0.5\textwidth]{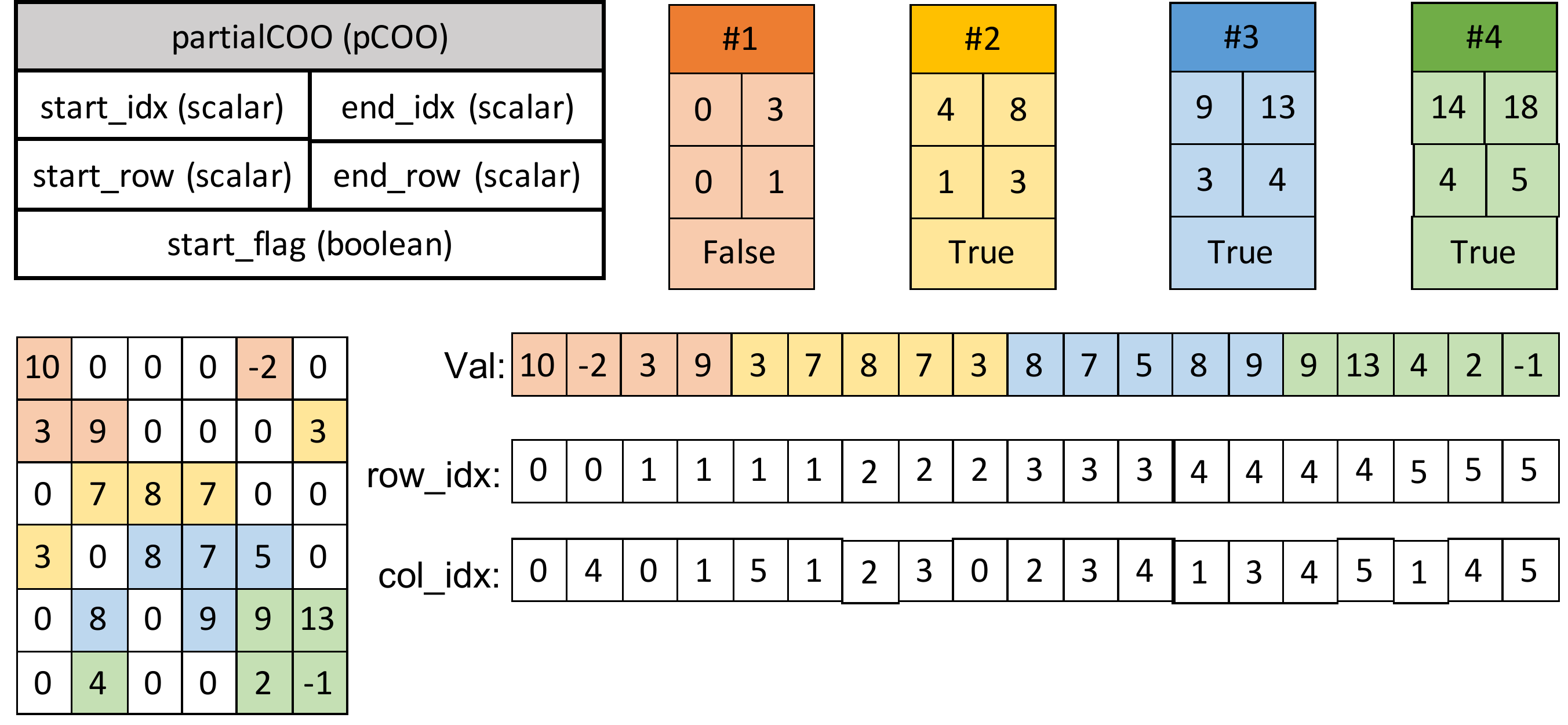}
    \caption{Example of partitioning a matrix into four parts using pCOO (sorted by row). }
    \label{pcoo-example}
\end{figure}

\begin{algorithm}
\caption{Launching COO-based SpMV kernel using pCOO}
\label{pcoo-kernel}
\KwIn{COO matrix A}
\KwIn{dense vector x ($n \times 1$)}
\KwIn{scalar $\alpha$ and $\beta$}
\KwInOut{dense vector y ($m \times 1$)}
\KwIn{Number of pCOOs: $np$}
\KwIn{pCOO matrix pA[np]}
Allocate space to hold partial results py[np]\\
\blue{/* Run in parallel on multi-GPU */} \\
\For{$i=1$ to $np$}{
    val = A.coo[pA[i].start\_idx] \\
    col\_idx = A.col\_idx[pA[i].start\_idx] \\
    row\_idx = A.row\_idx[pA[i].start\_idx] \\
    Launch: py[i]=<cooSpMVKernel>(val, col\_idx, row\_idx) \\
}
\For{$i=1$ to $np$}{
    \If{pA[i].start\_flag}{
         tmp = y[pA[i].start\_row] \\
    }
    memcpy: y[pA[i].start\_row] $\leftarrow$ py[i] \\
    \If{pA[i].start\_flag}{
        y[pA[i].start\_row] -= tmp * $\beta$ \\
    }
}

\end{algorithm}

\subsection{Managing SpMV Workload on Multi-GPUs}

\begin{figure}[]
    \centering
    \includegraphics[width=0.5\textwidth]{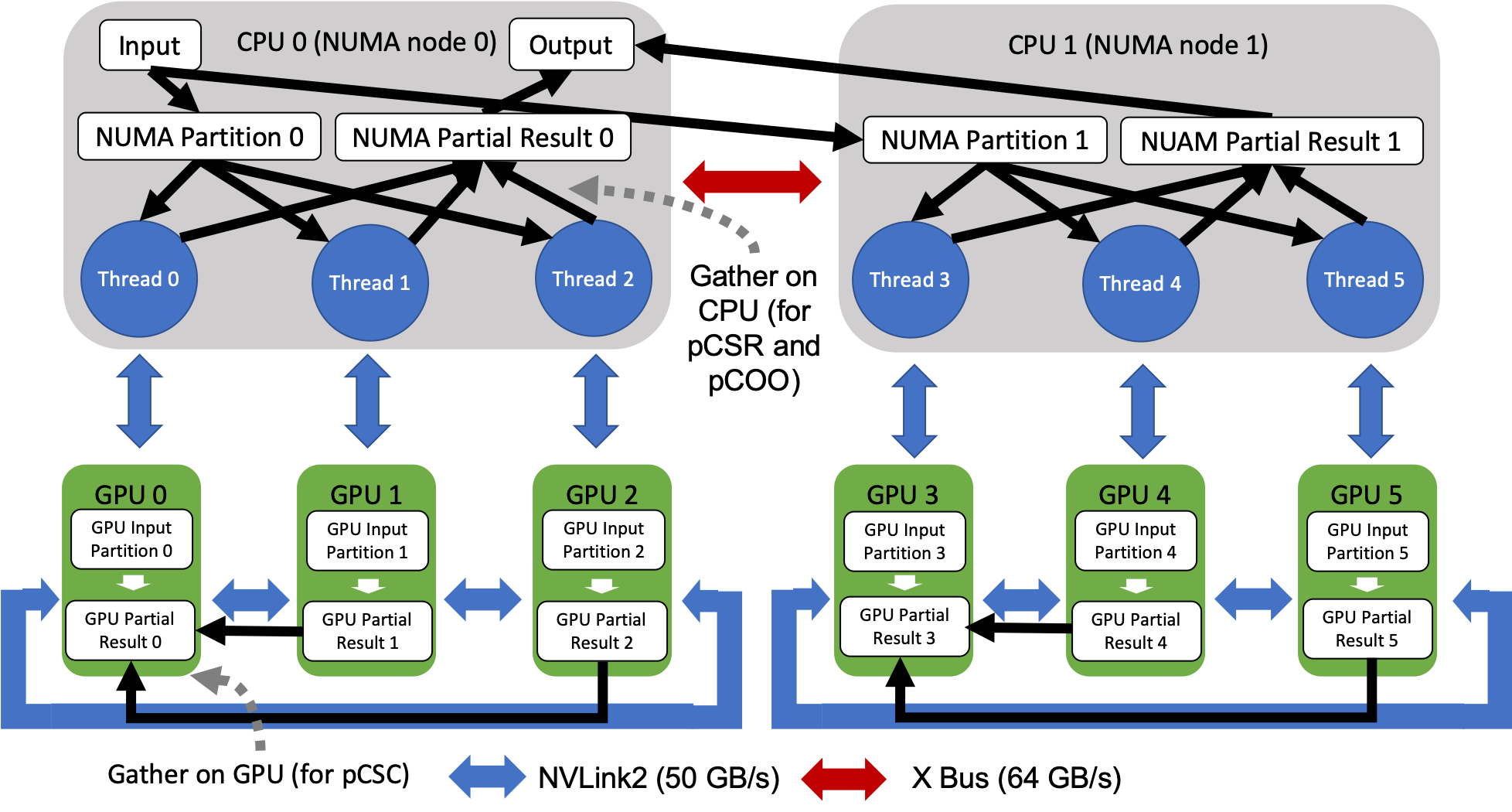}
    \caption{Managing SpMV workload on a compute node on Summit}
    \label{managing-smt}
\end{figure}

\begin{figure}[]
    \centering
    \includegraphics[width=0.5\textwidth]{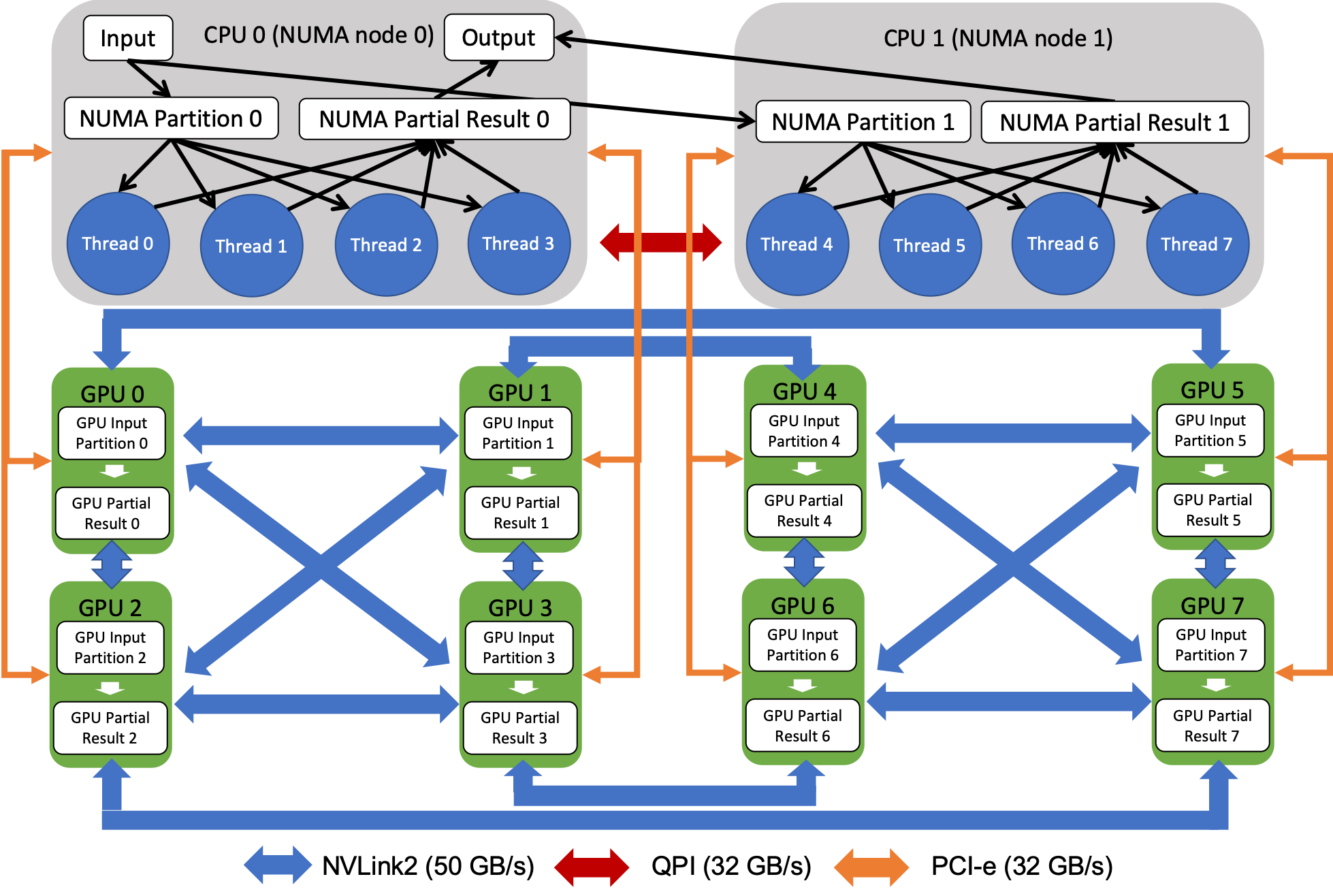}
    \caption{Managing SpMV workload on NVIDIA V100-DGX-1}
    \label{managing-dgx1}
\end{figure}

To efficiently manage multiple GPUs at the same time, we use one dedicated CPU thread to manage one GPU. Each thread is responsible for generating the workload partition for the corresponding GPU. If NUMA optimization (will be discussed in section \ref{numa-opt}) is enabled, threads on each NUMA node will also elect one representative thread to handle workload partitioning among NUMA nodes. Partial results will be gathered on CPUs for pCSR and pCOO or GPUs for pCSC. Details about partial result gathering will be discussed in section \ref{merg-opt}. \textbf{Fig. \ref{managing-smt}-\ref{managing-dgx1}} show how the SpMV workload is managed among the 6 GPUs on the computing node on Summit and 8 GPUs on the NVIDIA V100-DGX-1 system.

\section{Implementation Optimizations}
\label{opt}
In this section we cover several issues related to the implementations that are critical for achieving good performance and scalability.
\subsection{Workload partition}
\label{part-opt}
As will be seen in the evaluation, the workload partition can introduce considerable overhead up to 85\% on the tested matrices. Since each partition can be independently generated as discussed in Section~\ref{part}, we parallelize the partition process through multi-threading -- each thread is dedicated for a GPU. Additionally, we offload the most expensive workload to GPUs as specially designed kernels, e.g., the calculation of the local row pointers for CSR, column pointer for CSC, and row index array for COO. As data movement from CPU to GPUs is inevitable, this will not incur extra overhead.

\subsection{Handling NUMA Effect}
\label{numa-opt}
Since SpMV is a memory bound computation, the cost of data movement usually dominates the overall cost of the operation.
So, efficient copy of the partitioned workload represented by pCSR, pCSC, or pCOO to each of the GPU memory from the CPU memory is specially important. 
Dense GPU nodes usually partition the GPUs among several NUMA nodes. For example, on Summit six GPUs are partitioned among two NUMA nodes on each computing node as shown in \textbf{Fig. \ref{managing-smt}}. If the workload partitions are placed naively (i.e., on one NUMA node), then it is difficult to scale SpMV beyond three GPUs. This is mainly limited by both the CPU memory throughput within on NUMA node and the inter-connection speed in between NUMA nodes since GPUs on a different NUMA node need to fetch data through the inter-connection (e.g., X Bus on Summit).

\begin{figure}[]
    \centering
    \includegraphics[width=0.4\textwidth]{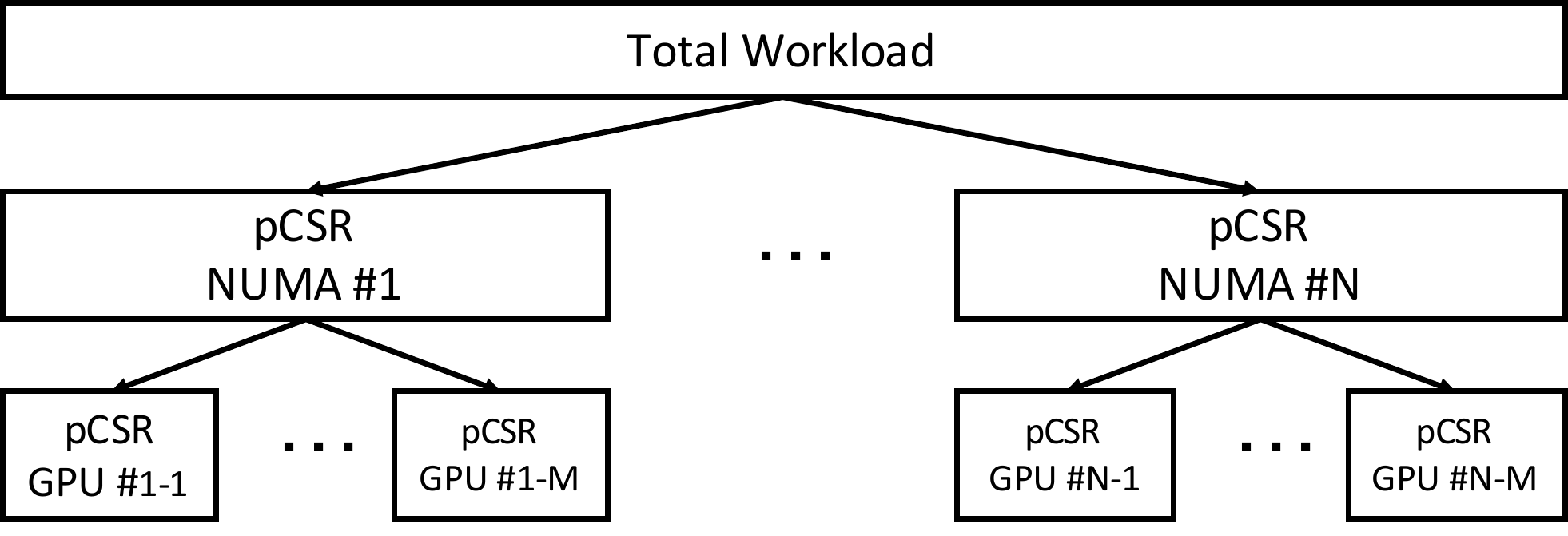}
    \vspace{-0.5em}
    \caption{Two Level Workload Partition.}
    
    \label{2level}
    \vspace{-0.5em}
\end{figure}

In this work, we design \texttt{mSpMV} to consider the NUMA effect to make sure pCSR, pCSC, or pCOO partitions are placed in among different NUMA nodes. The placement strategy is to place the number of workload partitions proportional to the number of GPUs on each NUMA node. This is done efficiently using our two level partitioning strategy. As shown in \textbf{Fig. \ref{2level}}, the first level partitions the workload among NUMA nodes and second level partition among GPUs. The two level partitioning strategy enables the work of partitioning itself parallelizaible. 

\subsection{Merging Partial Results}
\label{merg-opt}

\begin{figure}[]
    \centering
    \includegraphics[width=0.4\textwidth]{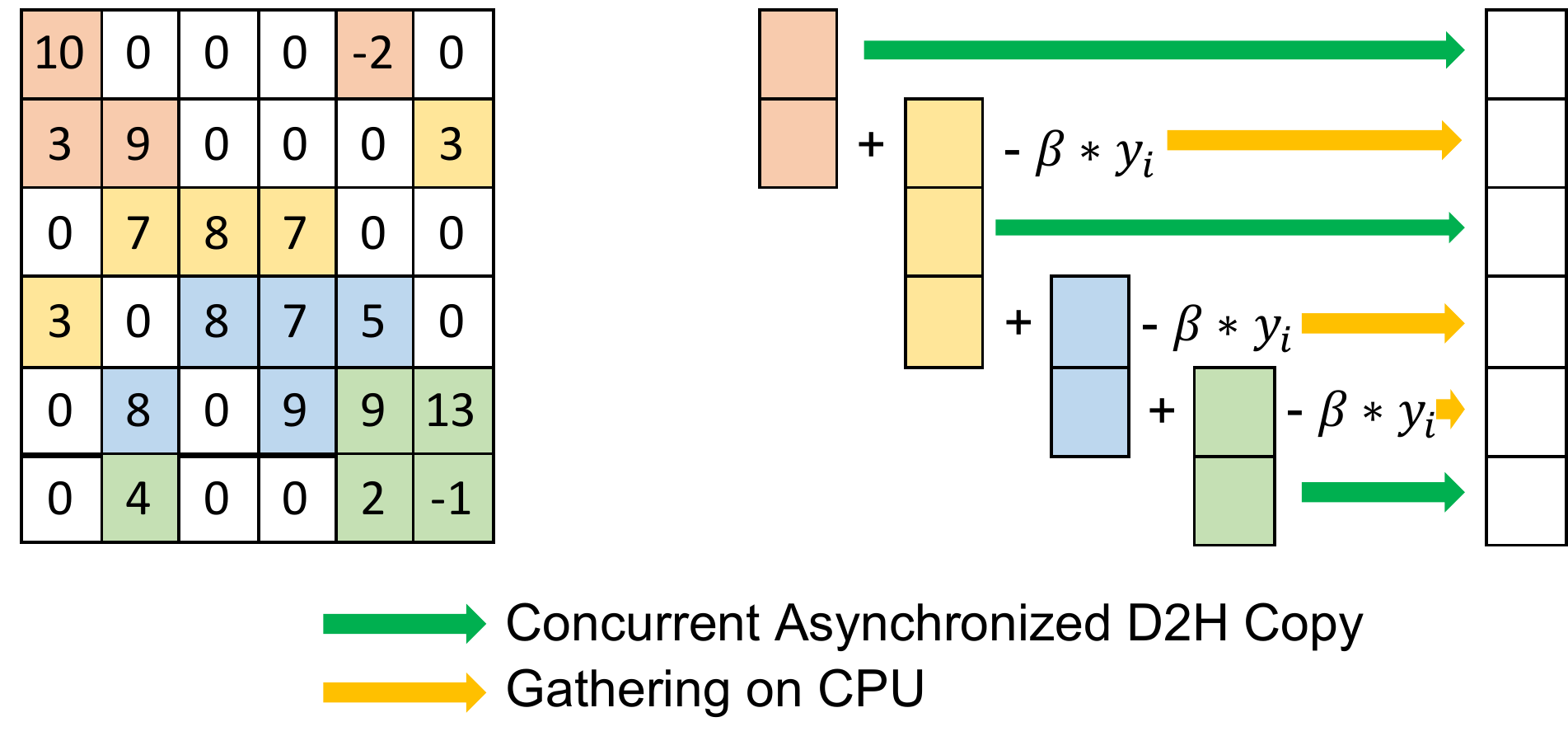}
    \vspace{-0.5em}
    \caption{Result Merging for pCSR and pCOO (sorted by row).}
    \label{row-gather}
    \vspace{-0.5em}
\end{figure}

\begin{figure}[]
    \centering
    \includegraphics[width=0.4\textwidth]{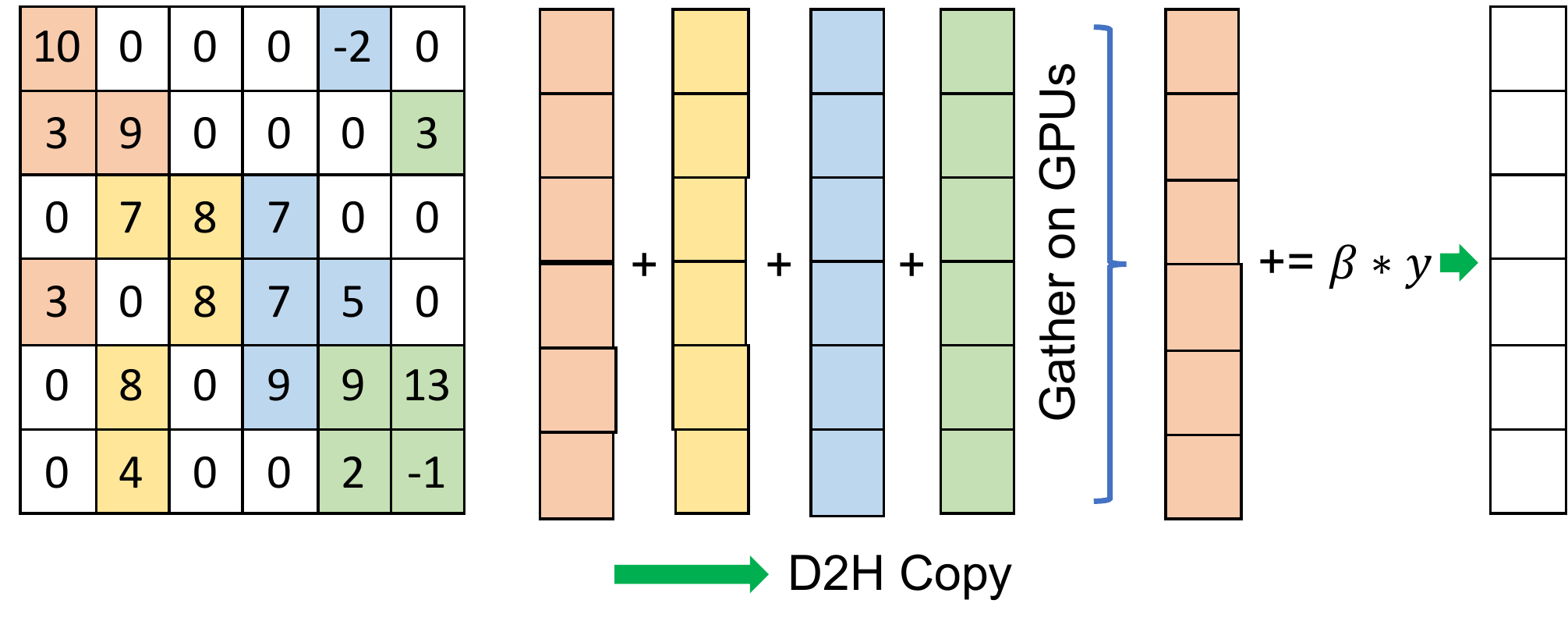}
    \caption{Result Merging for pCSC and pCOO (sorted by column).}
    \label{col-gather}
     \vspace{-0.5em}
\end{figure}

Another issue that can impact the of performance of SpMV on multi-GPU system is how partial results are merged into the final result. The partitioning format used can impact how the partial results can be merged. Basically they can be classified into two categories: row-based partitioning and column-based partitioning. 

Row-based partitioning such as pCSR and pCOO (sorted by row) assigns consecutive rows to a single partition. So, the result of each partition is a segment of the final result vector except for the element at each end, which may be partial result if the current partition share same rows with others. To optimize the merging process, we use GPU-CPU copy to directly copy the non-overlapping result to the final position on the CPU memory and let CPU handle overlapping elements. It brings relative low performance impact since (1) Memory copy can be done concurrently; (2) Since the overlapping issue only need to be handled $np$ times.


Column-based partitioning such as pCSC and pCOO (sorted by column) assigns consecutive columns to a single partition. So, the result of each partition is a vector with the same dimension as the final result but each element in the vector is partial result. To optimize the merging process, we first let all GPUs gather their partial results to one GPU and then copy the result back to CPU.
\section{Experimental Evaluation}
\label{experiements}
In this section, we report our experimental results and discuss our observations in detail.
\subsection{Evaluation Platform}
We evaluate our framework on two multi-GPU platforms: the Summit supercomputer at Oak Ridge National Laboratory and an NVIDIA V100-DGX-1 system. On Summit, one computing node is used for computation. Each node is equipped with 6 Nvidia Tesla V100 GPUs with 16 GB memory on each GPU and two IBM POWER9 CPUs with 512 GB memory. GPU-GPU and CPU-GPU are inter-connected via NVLinks. CPU-CPU are inter-connected via X-Bus. On the DGX-1 system, the whole system is used for computation. The DGX-1 system is equipped with 8 Nvidia Tesla V100 GPUs with 16 GB memory and two 20-Core Intel Xeon
E5-2698 v4 with 512 GB memory. GPUs are inter-connected via NVLinks. CPU-GPU are inter-connected via PCIe. CPUs are inter-connected via QPI.

We compiled our framework with GCC 7.4.0 and CUDA 10.1.168. We employ OpenMP (version 4.5) for spawning tasks on the GPUs
For single-GPU kernel execution, we utilize the CSR-based SpMV kernel in cuSparse. If inputs are in CSC format, the kernel is invoked with transpose on so as to avoid format conversion. For COO based inputs, a GPU-based COO-to-CSR conversion kernel is invoked first before the SpMV kernel. We include the execution time of both conversion step and  computation step in our reported time. 

\subsection{Selected Matrices}
\textbf{Table \ref{input-matrices}} lists the set of sparse matrices we use in our evaluation, collected from the SuitSparse Matrix Collection~\cite{sparse-matrix-collection}. The matrices are ordered by the number of non-zero elements. When selecting the matrices, we choose matrices with strong power-law characteristics (skewed degree distribution)~\cite{newman2005power}. Such pattern is commonly observed in social networks and web graphs. 
As reported by Yang et al.~\cite{Yang2011}, the number of non-zeros in the columns of these matrices follow a power-law distribution.
Hence, we apply this rule to locate these matrices.
The distribution in the form: $P(k) \sim k^{-R}$ is considered to follow power law. Here, $R$ is called the exponent of the power law and is calculated from the distribution of non-zero elements, $k$ denotes the number of non-zero elements per column. Usually, choosing a value for $R$ within the interval $[1, 4]$  correlates to strong phenomenon of power law~\cite{li2013smat}.

\begin{table}[]
\small
\caption{List of sparse matrices used for evaluation.}
\label{input-matrices}
\begin{tabular}{|c|c|c|c|}
\hline
Matrix & row $\times$ col & nnz & R \\ \hline
mouse\_gene &45K $\times$ 45K         & 28M    &    1.03    \\ \hline
wb-edu &9M $\times$ 9M         & 57M    &   2.13     \\ \hline
com-LiveJournal &3M $\times$ 3M         & 69M    &   2.40      \\ \hline
hollywood-2009 &1M $\times$ 1M         & 113M    &  1.92     \\ \hline
com-Orkut & 3M $\times$ 3M        &  234M   &  2.13   \\ \hline
HV15R     & 2M $\times$ 2M        &  283M   &    3.09    \\ \hline
\end{tabular}
\vspace{-1em}
\end{table}

\subsection{Evaluation Methodology}
To evaluate our multi-GPU SpMV design, we implement the following versions:
\begin{itemize}[leftmargin=*]
    
\item \textbf{Baseline} is a simple multi-GPU SpMV that we consider as the baseline for comparison purpose. In the baseline version, the input matrix is partitioned 
in either row blocks (for CSR and COO) or column blocks (for CSC) without considering the distribution of non-zero elements. Moreover, no multi-threading is involved for partitioning the workload or managing the GPUs. Both workload partitioning and partial result merging are done on the CPU without optimization. In addition, no NUMA-aware optimization has been applied.

\item \textbf{p*} refers to the implementation that leverages our pCSR, pCSC, and pCOO datastructures to achieve workload balance. Multi-threading is used for partitioning the workload, merging the result, and managing  the GPUs in parallel. However, no optimization is performed.

\item \textbf{p*-opt} is based on p*. All implementation optimizations are applied to this implementation.

\end{itemize}

\begin{figure*}[t]
    \centering
    \begin{subfigure}[t]{0.49\textwidth}
    \includegraphics[width=\textwidth]{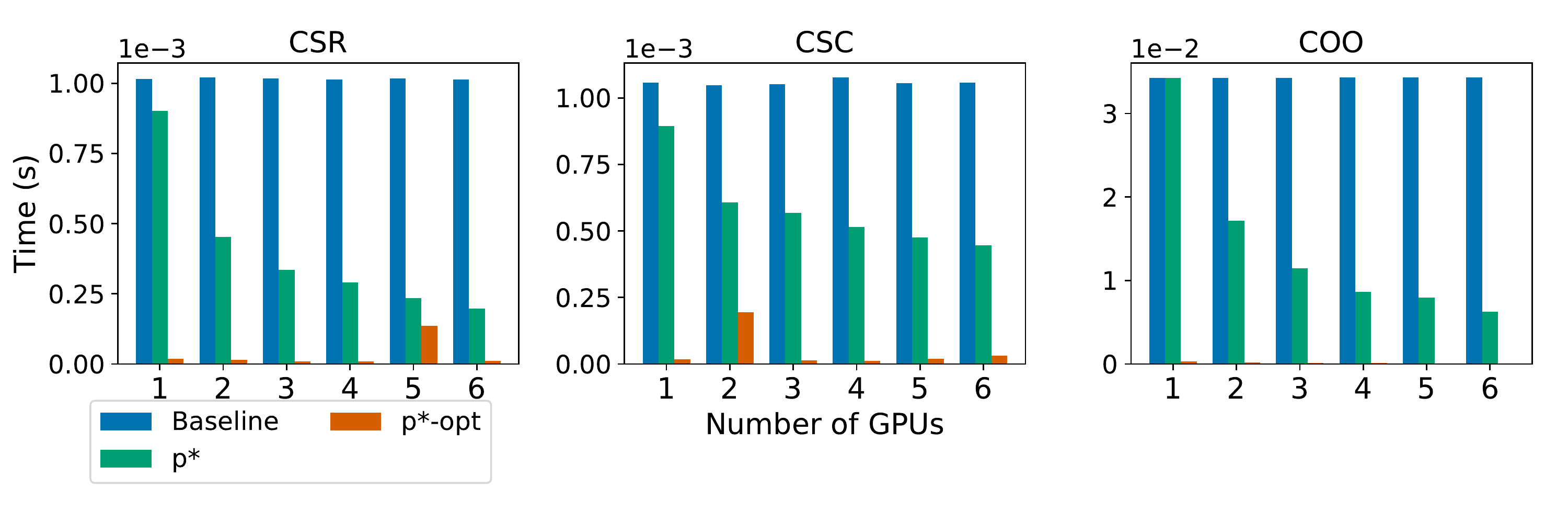}
    \vspace{-2em}
    \caption{ORNL Summit}
    \end{subfigure}
    \begin{subfigure}[t]{0.49\textwidth}
    \includegraphics[width=\textwidth]{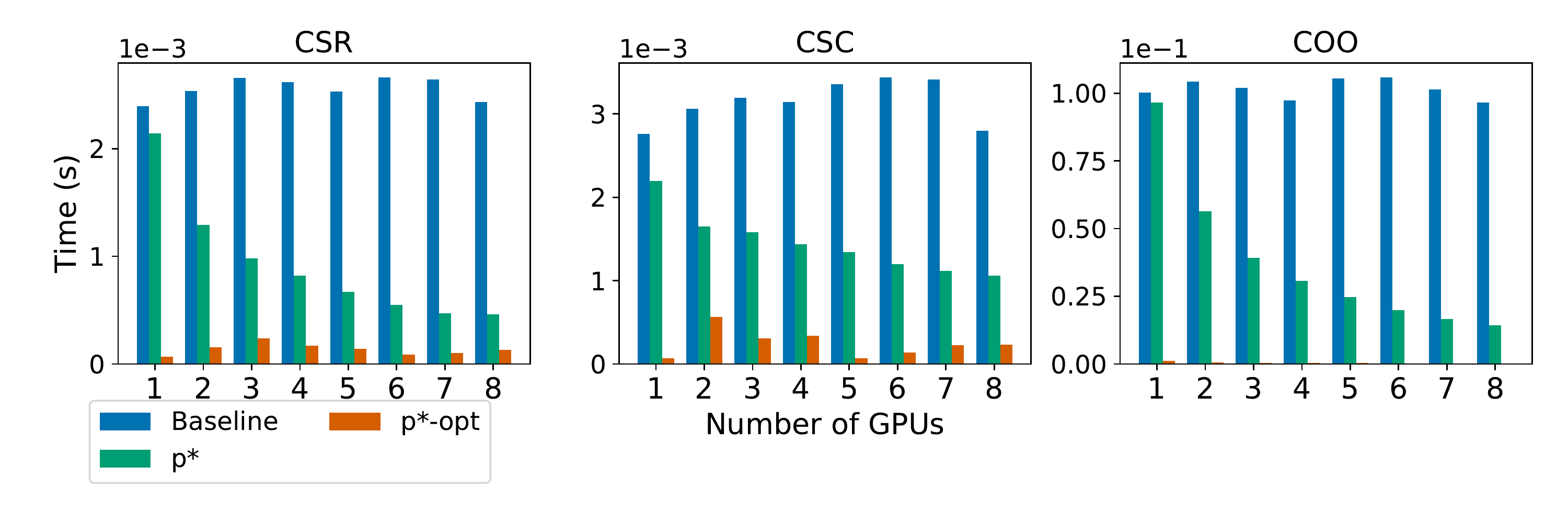}
    \vspace{-2em}
    \caption{NVIDIA V100-DGX-1}
    \end{subfigure}
    \caption{Time cost of workload partition (input matrix: HV15R)}
    \label{part-time}
    \vspace{-1em}
\end{figure*}

\begin{figure*}[t]
    \centering
    \begin{subfigure}[t]{0.49\textwidth}
    \includegraphics[width=\textwidth]{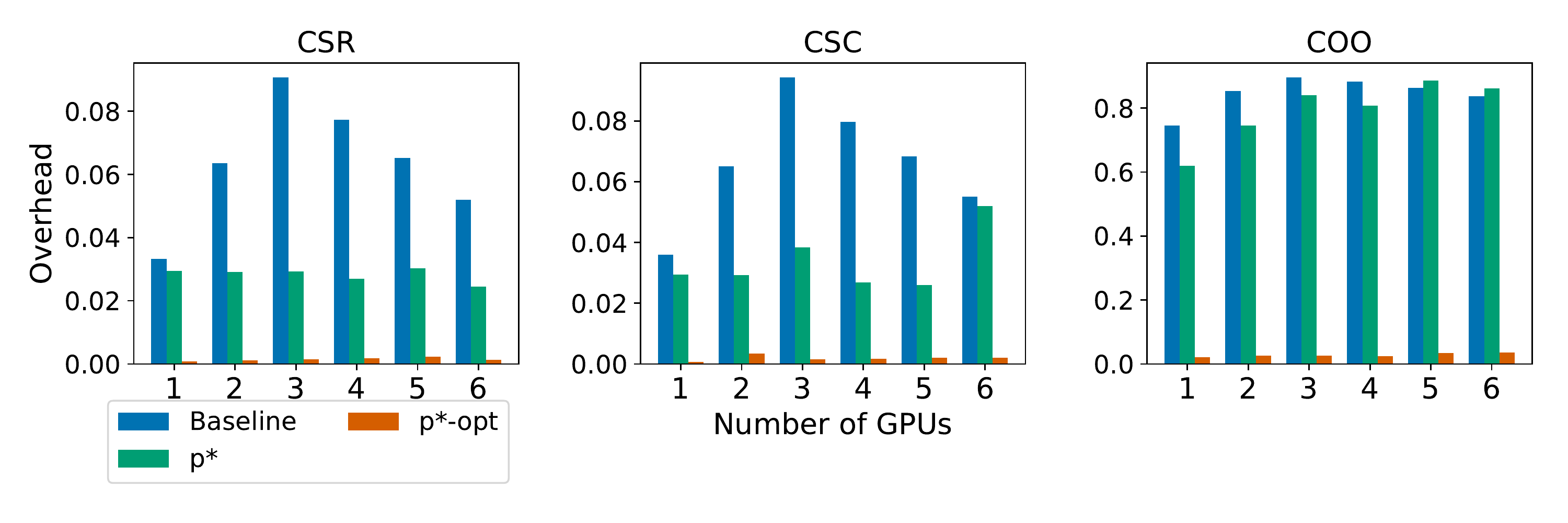}
    \vspace{-2em}
    \caption{ORNL Summit}
    \end{subfigure}
    \begin{subfigure}[t]{0.49\textwidth}
    \includegraphics[width=\textwidth]{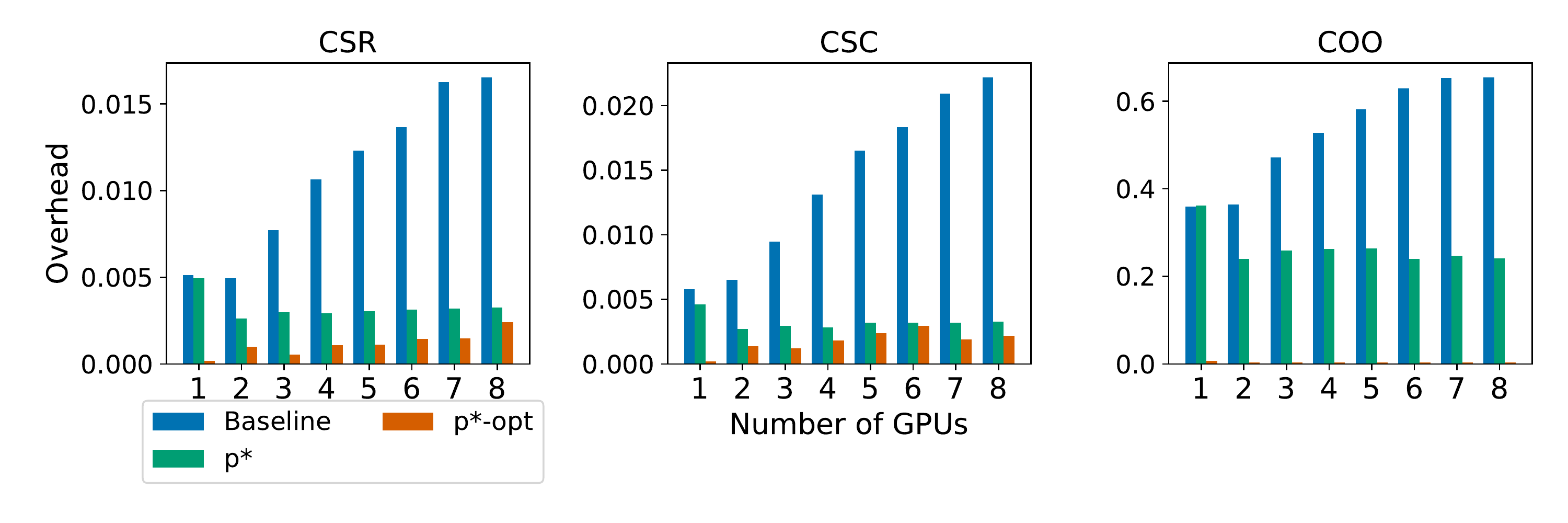}
    \vspace{-2em}
    \caption{NVIDIA V100-DGX-1}
    \end{subfigure}
    \caption{Overhead workload partition (input matrix: HV15R)}
    \label{part-overhead}
    \vspace{-1em}
\end{figure*}

\begin{figure*}[t]
    \footnotesize
    \begin{subfigure}[t]{0.49\textwidth}
    \includegraphics[width=\textwidth]{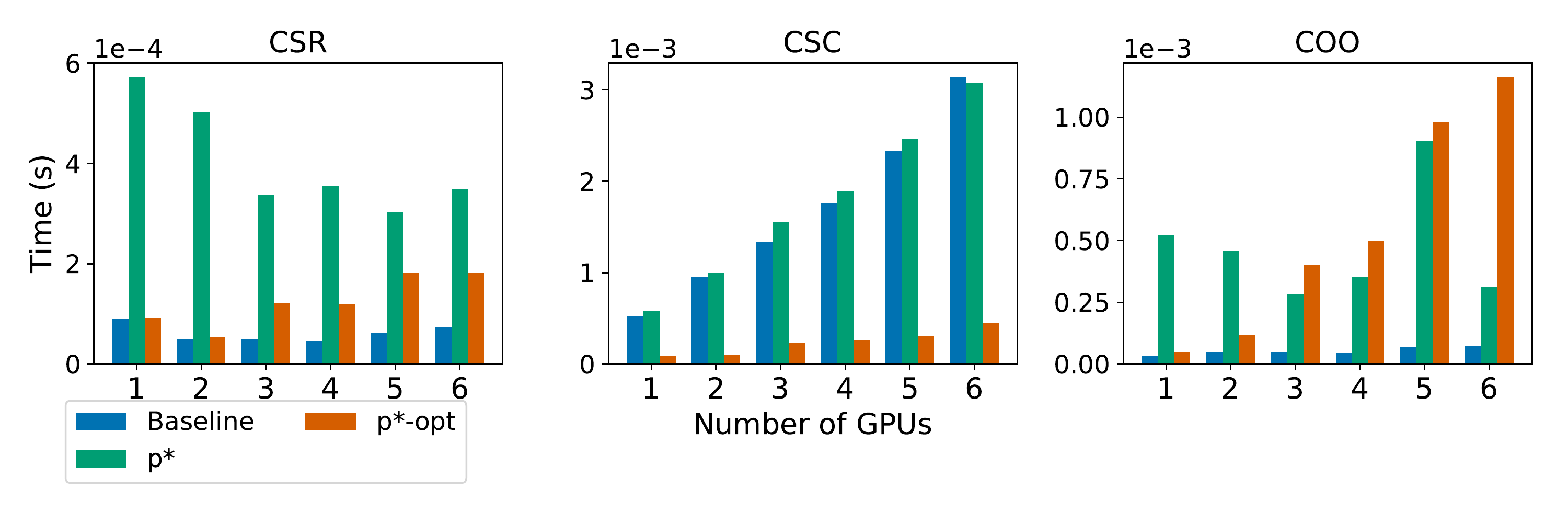}
    \vspace{-2em}
    \caption{ORNL Summit}
    \end{subfigure}
    \begin{subfigure}[t]{0.49\textwidth}
    \includegraphics[width=\textwidth]{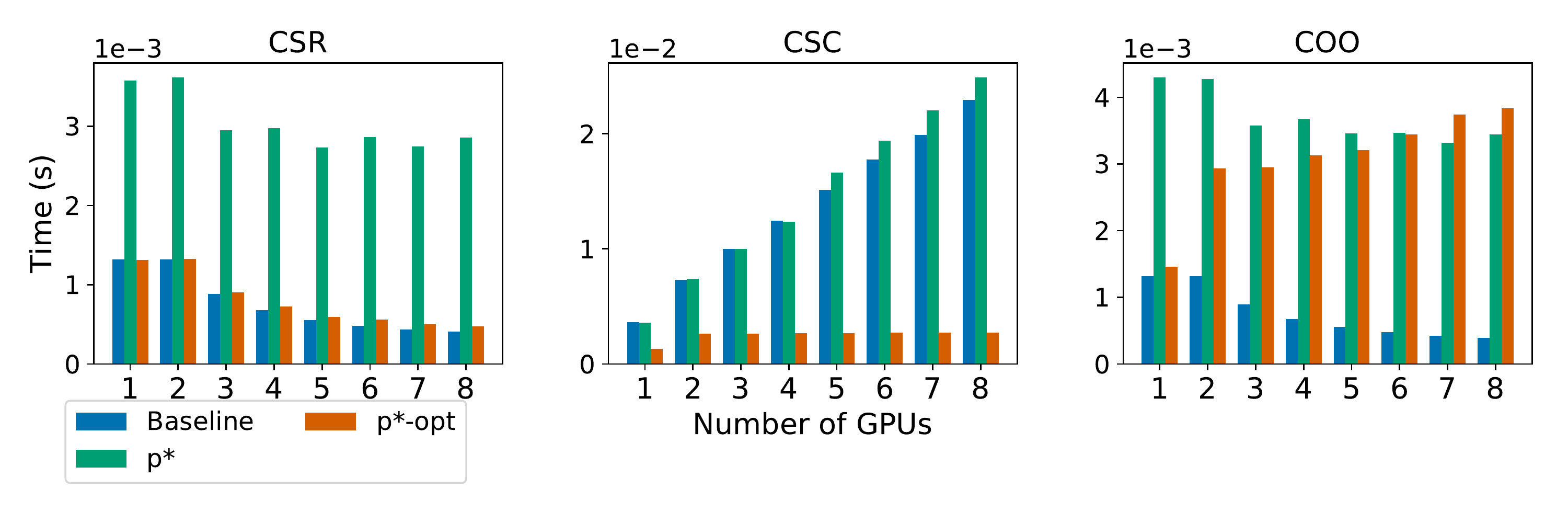}
    \vspace{-2em}
    \caption{NVIDIA V100-DGX-1}
    \end{subfigure}
    \caption{Time cost of merging partial results (input matrix: HV15R).}
    \vspace{-1em}
    
    \label{merg-time}
\end{figure*}

\begin{figure*}[t]
    \centering
    \begin{subfigure}[t]{0.49\textwidth}
    \includegraphics[width=\textwidth]{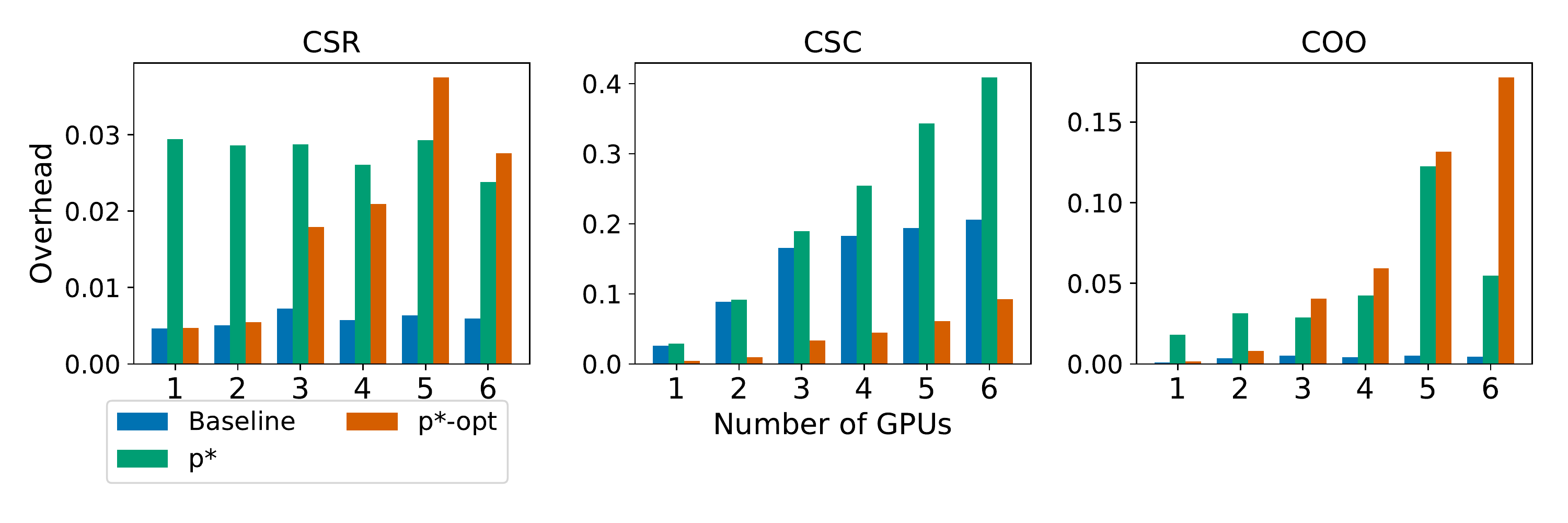}
    \vspace{-2em}
    \caption{ORNL Summit}
    \end{subfigure}
    \begin{subfigure}[t]{0.49\textwidth}
    \includegraphics[width=\textwidth]{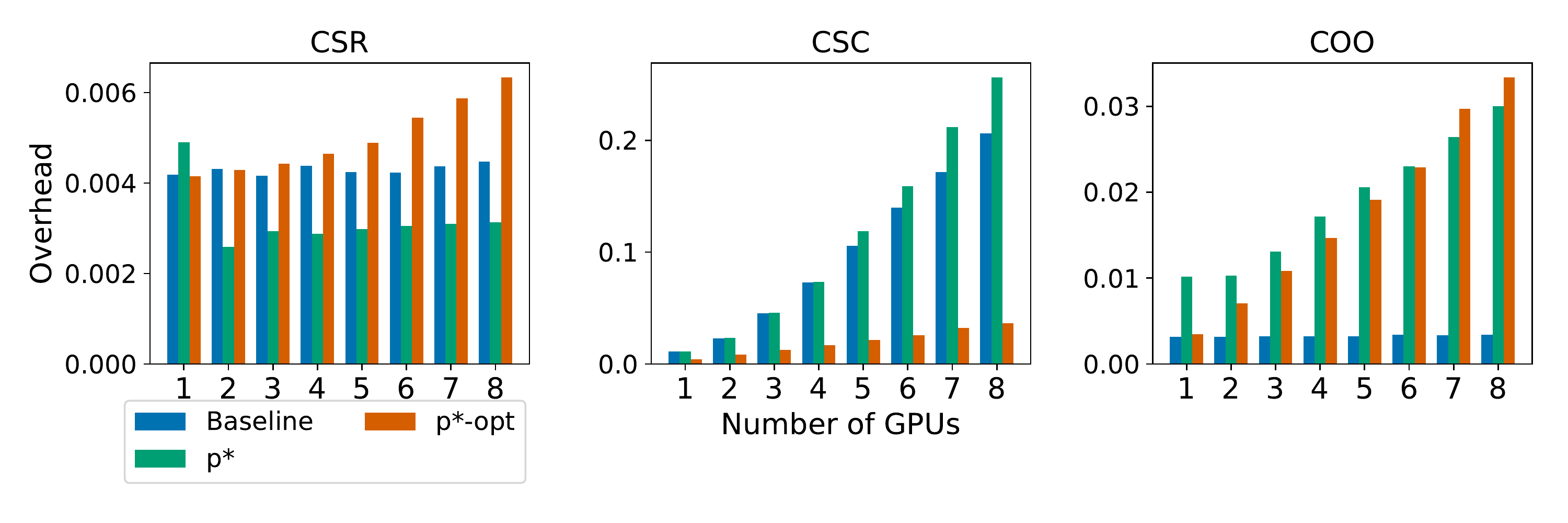}
    \vspace{-2em}
    \caption{NVIDIA V100-DGX-1}
    \end{subfigure}
    \caption{Overhead of merging partial results (input matrix: HV15R).}
    \vspace{-1em}
    \label{merg-time}
\end{figure*}

\subsection{Partitioning Overhead}
We report partitioning overhead as the percentage of total execution time spent in partitioning the input matrix and distributing the partitions to the GPUs.
\textbf{Fig. \ref{part-time}} shows the overhead of workload partitioning on Summit and DGX-1. 
In the baseline implementation, the input matrix is partitioned into row/column blocks.
With CSR and CSC format, the main cost comes from calculating the local row/column pointers. However, calculating row/column pointers takes only constant time ($O(m)$ or $O(n)$). 
If the size of the matrix is large, partitioning it on CPUs can be time consuming. 
For example, large matrix partitioning incurs 3.8\% - 9\% overhead on Summit and 0.5\%-2.3\% on DGX-1. 
For COO, the partitioning is done by searching  the row pointer array and finding the start and end positions of each row block. 
The most expensive part for partitioning COO comes from calculating the right row index array. Compared to the CSR and CSC representation, this operation is more expensive ($O(nnz)$ compared to $O(m)$ or $O(n)$ for CSR or CSC). 
From Fig. \ref{part-time}, we observe 72\% - 85\% overhead on Summit and 38\% - 62\% on DGX-1 for COO partitioning.

Compared to the baseline partitioning, our pCSR, pCSC, and pCOO (shown as p*) do not incur any extra partition overhead. 
By parallelizing the partitioning process, we achieve good speedup with increasing  number of GPUs.
However, partitioning these data structures still encounter high overhead due to the involvement of the CPUs.

On the other hand, from \textbf{Fig. \ref{part-time}}, we can see that, p*-opt gains significant benefit by offloading time consuming part on to the GPUs as mentioned in section \ref{part-opt} and thus reducing the overhead.
By applying this optimization, partitioning overhead was reduced to less than 2\% for most cases.
 
\begin{figure*}[t]
    \centering
    \begin{subfigure}[t]{0.49\textwidth}
    \includegraphics[width=\textwidth]{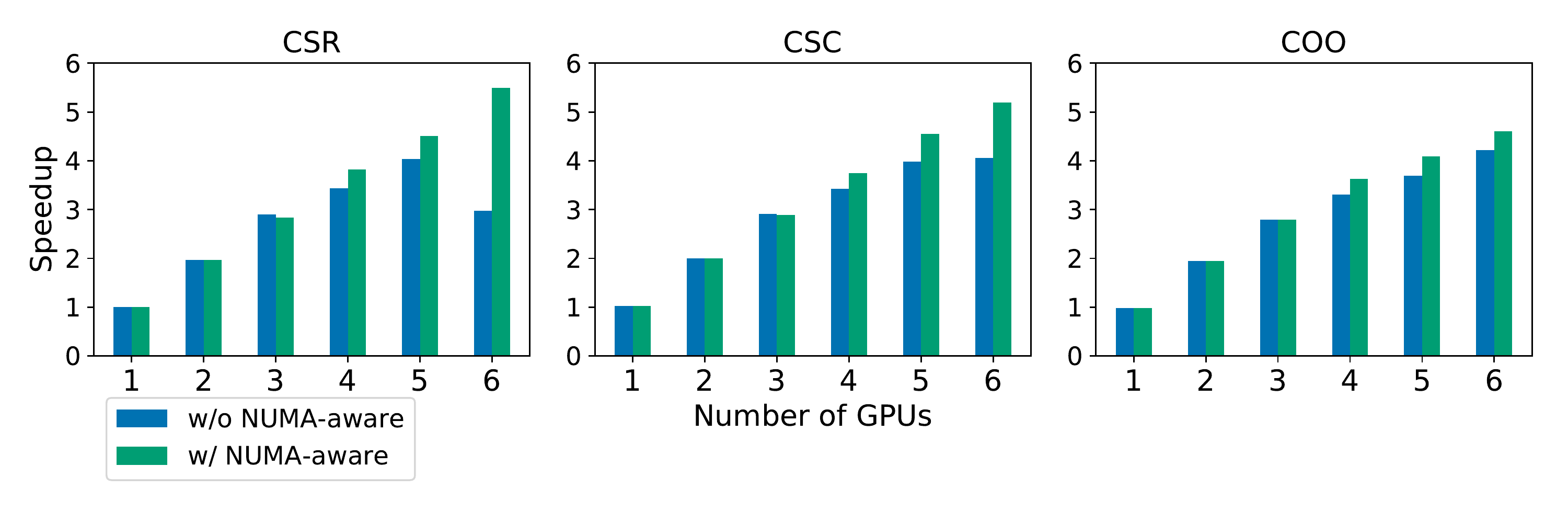}
    \vspace{-2em}
    \caption{ORNL Summit}
    \end{subfigure}
    \begin{subfigure}[t]{0.49\textwidth}
    \includegraphics[width=\textwidth]{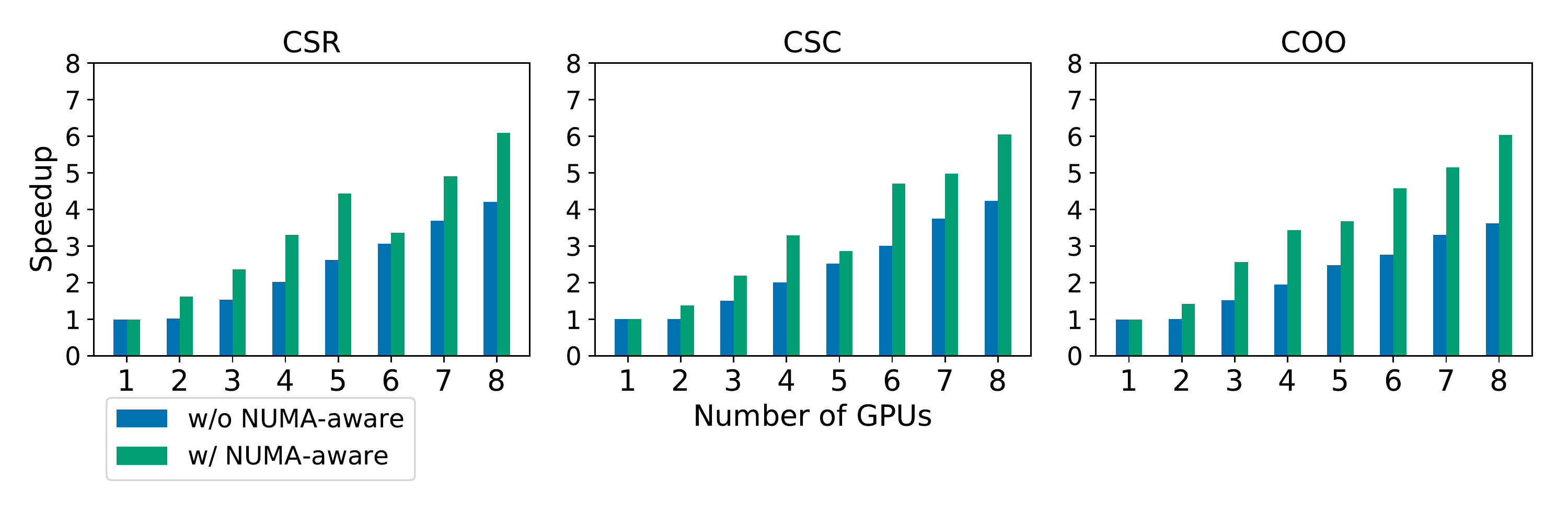}
    \vspace{-2em}
    \caption{NVIDIA V100-DGX-1}
    \end{subfigure}
    \caption{Comparing speedup with and without NUMA-aware (input matrix: HV15R)}
    \label{numa-speedup}
    \vspace{-1em}
\end{figure*}

\begin{figure*}[t]
    \centering
    \begin{subfigure}[t]{0.49\textwidth}
    \includegraphics[width=\textwidth]{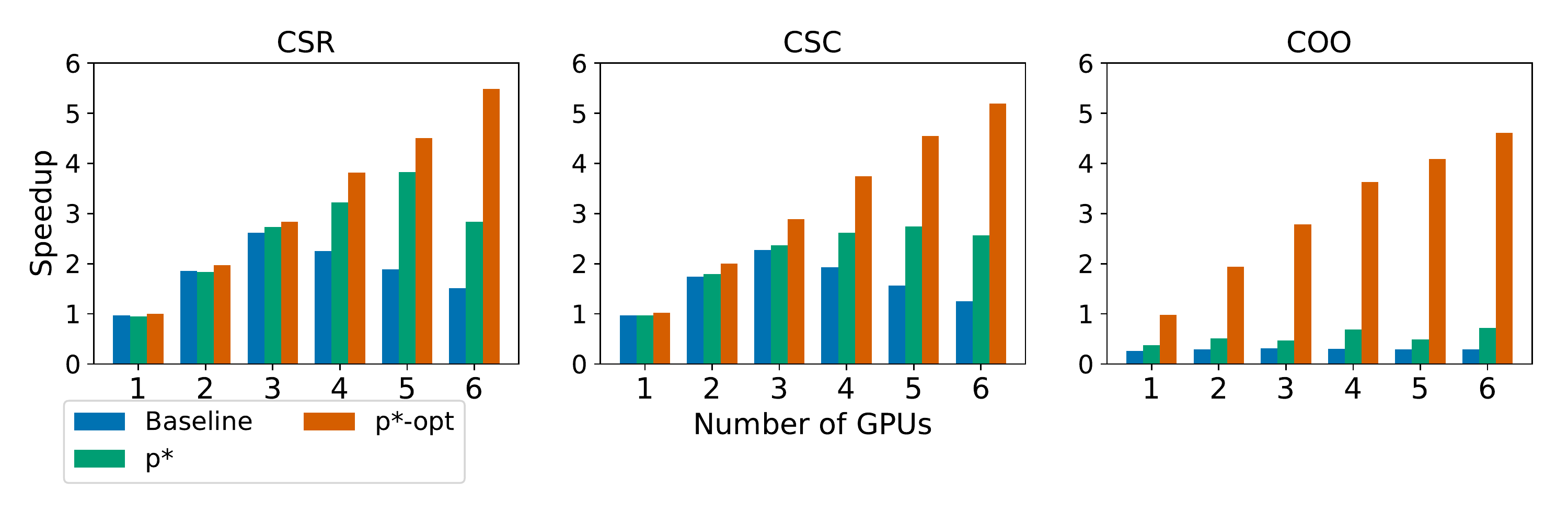}
    \vspace{-2em}
    \caption{ORNL Summit}
    \end{subfigure}
    \begin{subfigure}[t]{0.49\textwidth}
    \includegraphics[width=\textwidth]{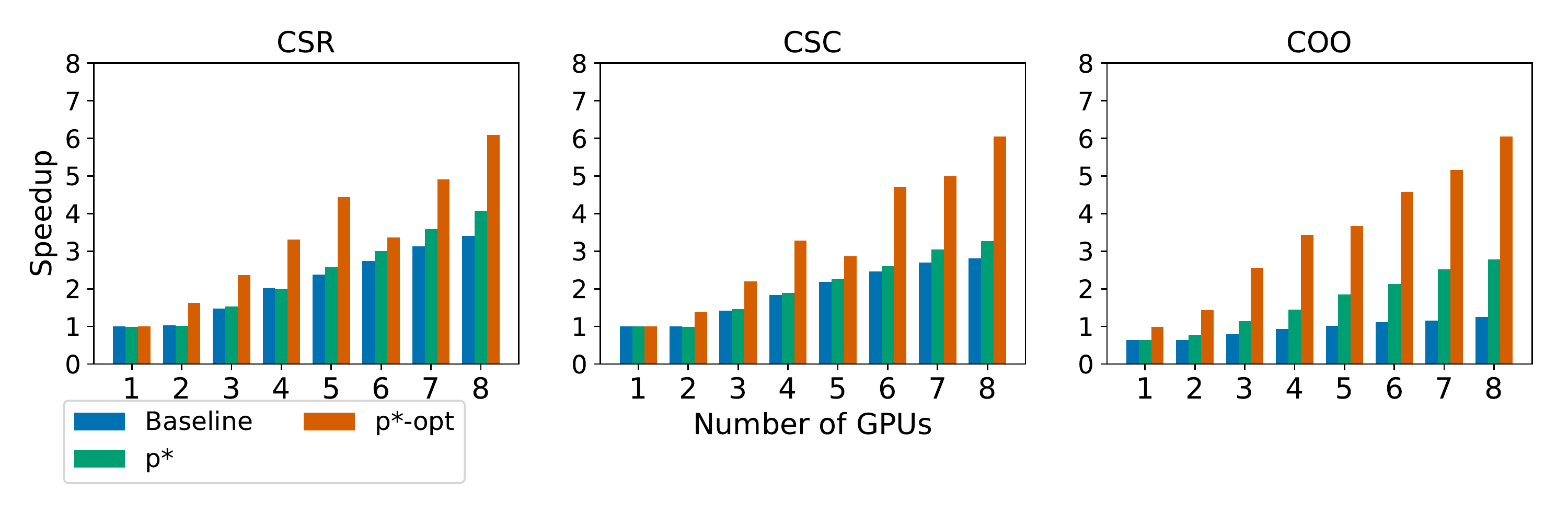}
    \vspace{-2em}
    \caption{NVIDIA V100-DGX-1}
    \end{subfigure}
    \caption{Comparing total speedup (input matrix: HV15R).}
    \label{total-speedup}
    \vspace{-1em}
\end{figure*}

\begin{figure*}[t]
    \centering
    \begin{subfigure}[t]{0.33\textwidth}
    \includegraphics[width=\textwidth]{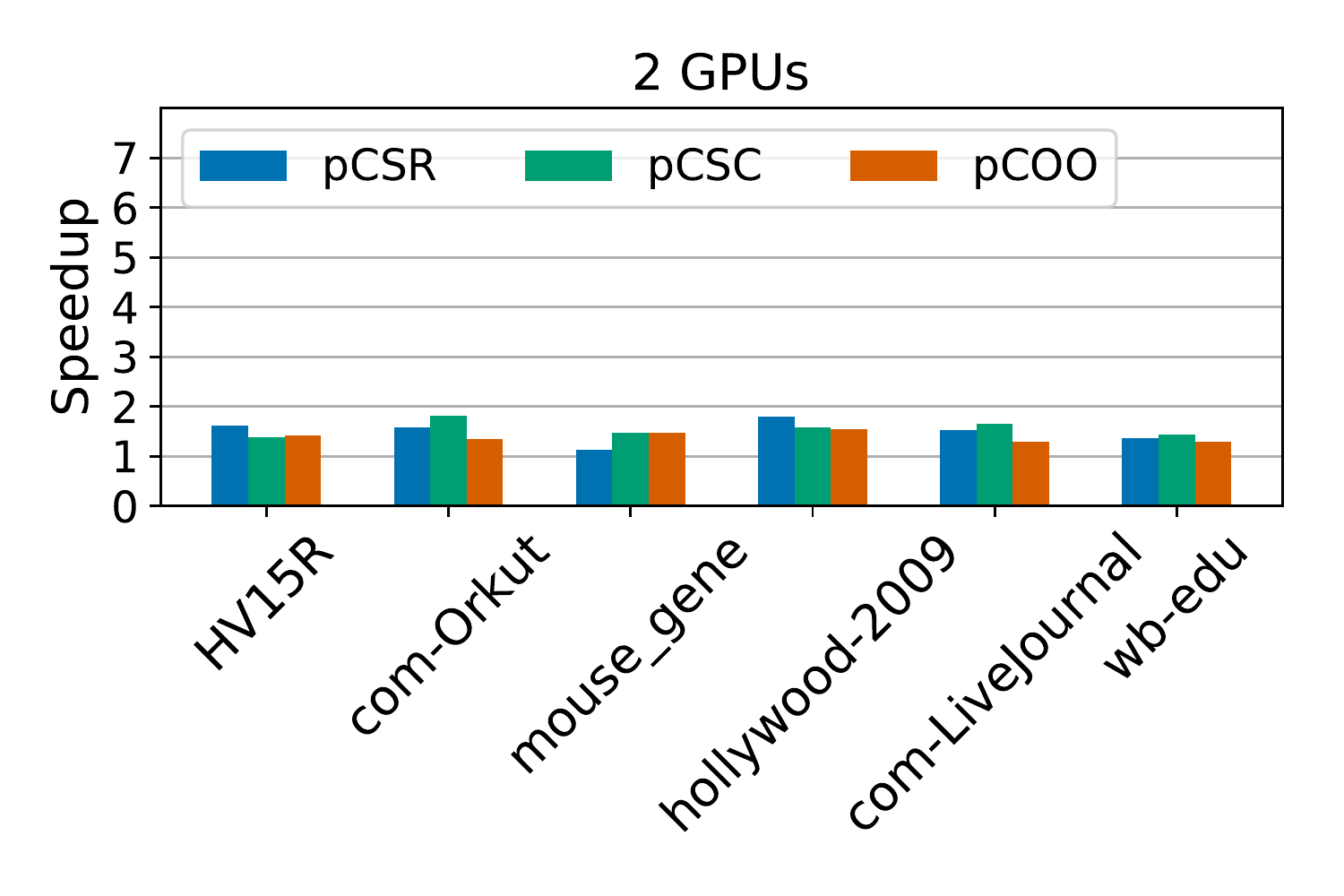}
    \end{subfigure}
    \begin{subfigure}[t]{0.33\textwidth}
    \includegraphics[width=\textwidth]{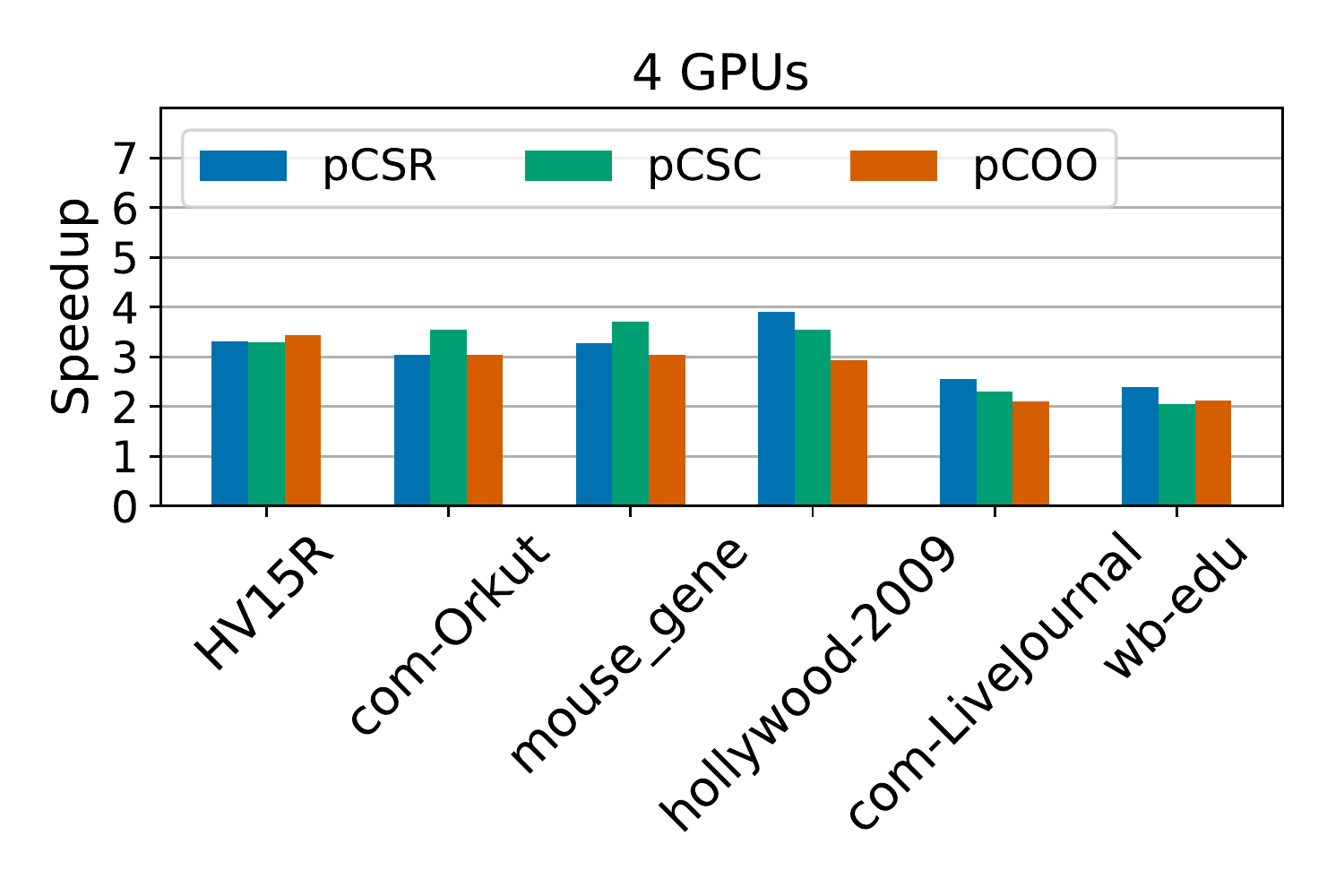}
    \end{subfigure}
    \begin{subfigure}[t]{0.33\textwidth}
    \includegraphics[width=\textwidth]{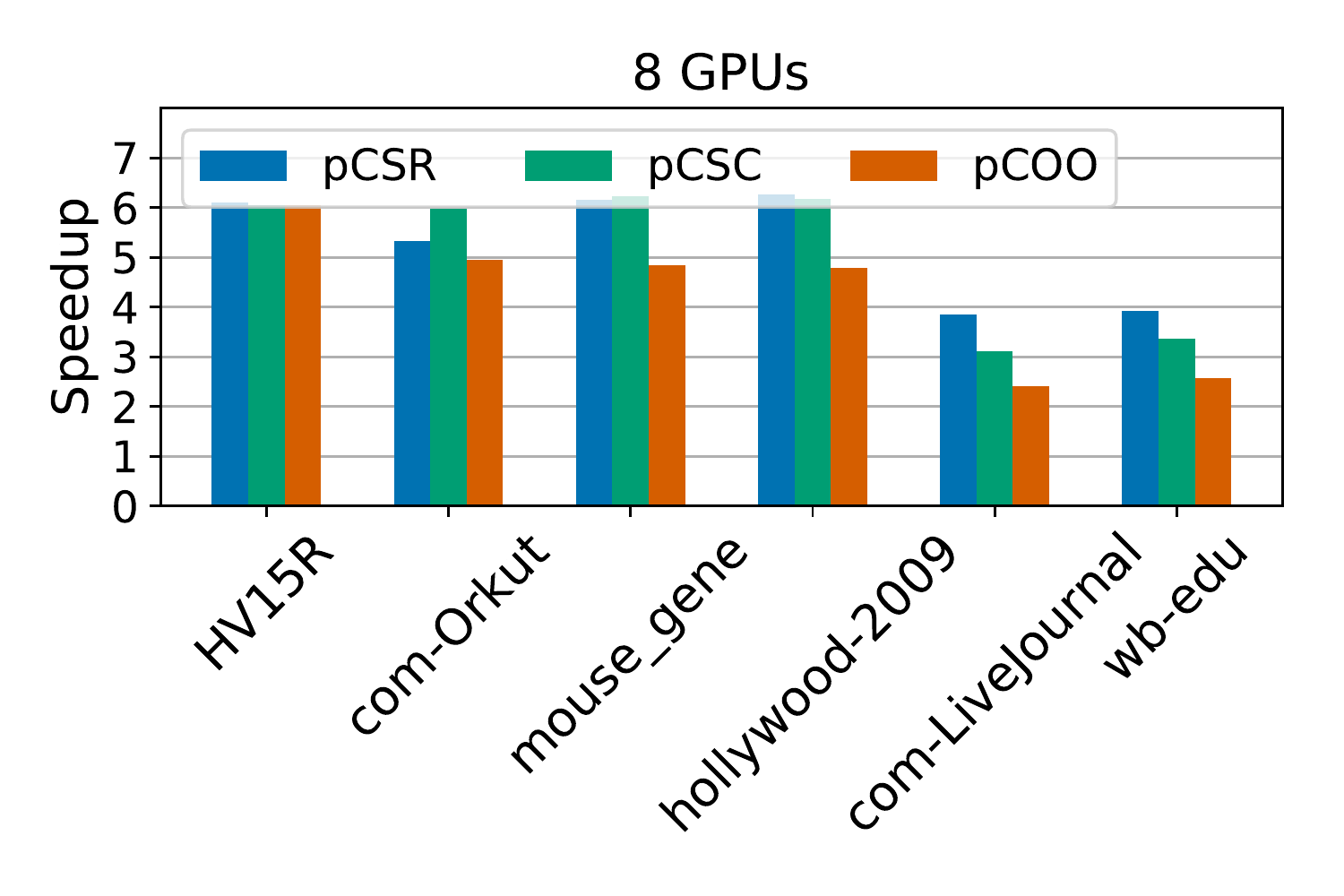}
    \end{subfigure}
    \vspace{-2em}
    \caption{Speedup comparison using different matrices on NVIDIA V100-DGX-1.}
    \label{merg-time}
     \vspace{-1em}
    \label{multi-dgx1}
\end{figure*}

\begin{figure*}[t]
    \centering
    \begin{subfigure}[t]{0.33\textwidth}
    \includegraphics[width=\textwidth]{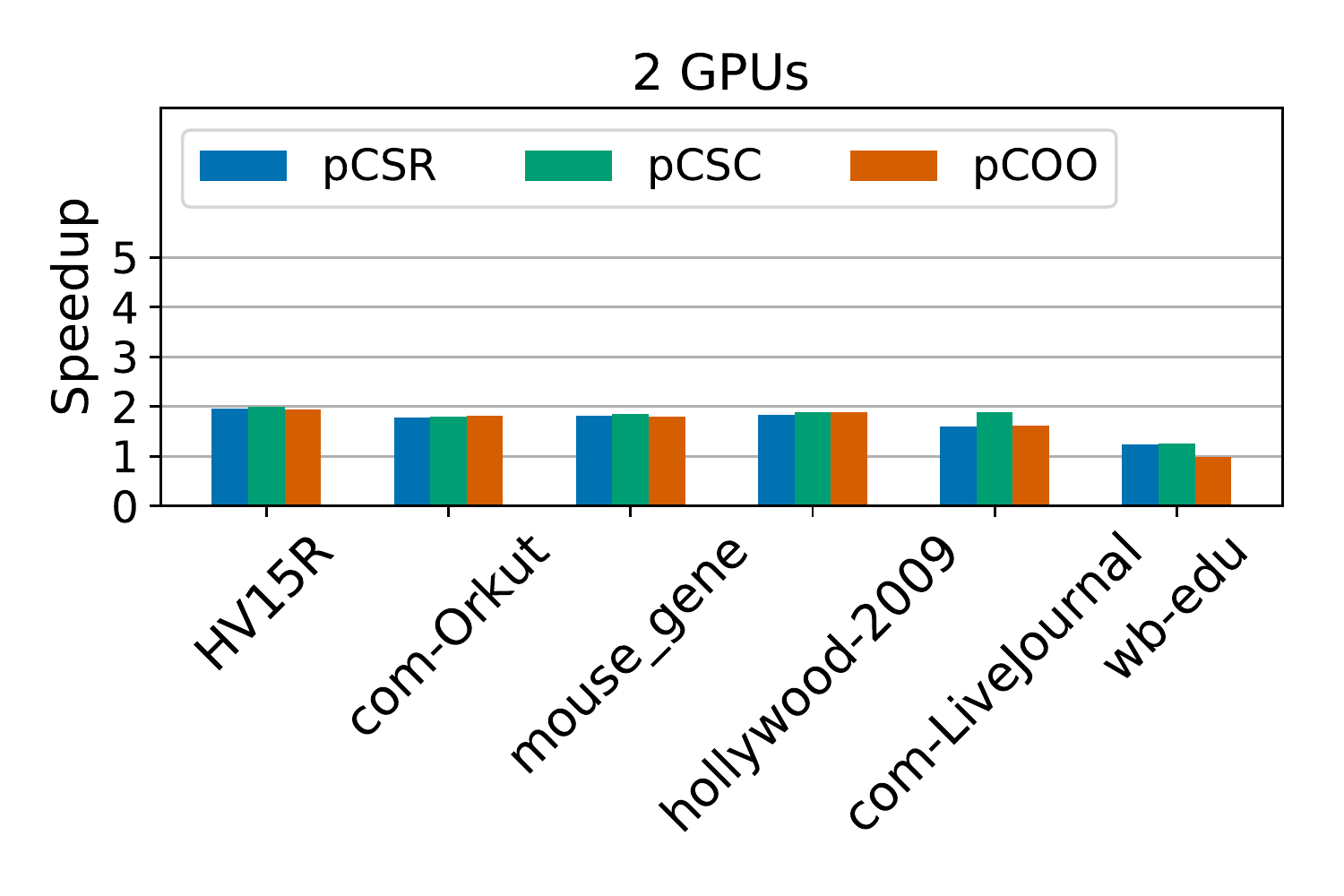}
    \end{subfigure}
    \begin{subfigure}[t]{0.33\textwidth}
    \includegraphics[width=\textwidth]{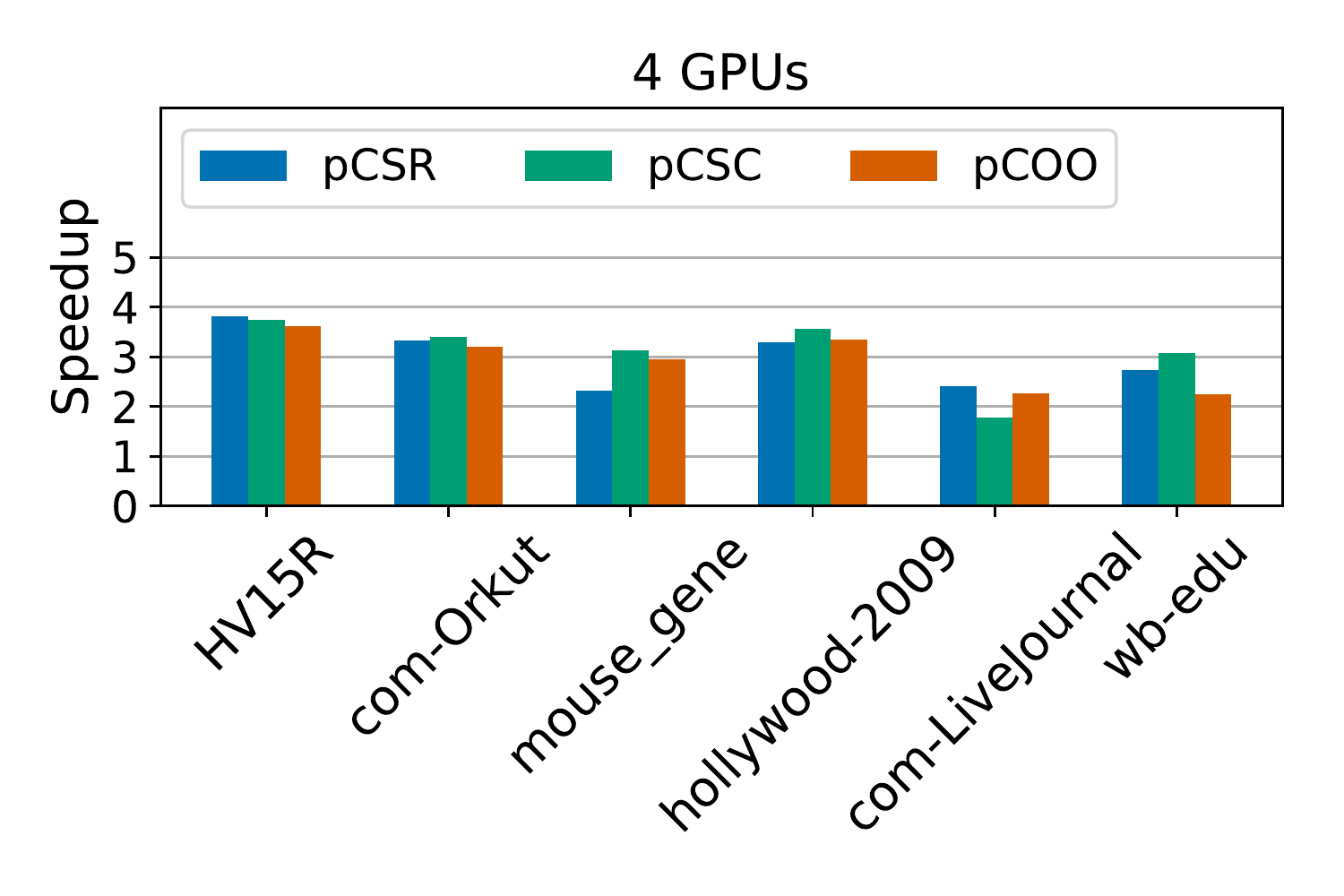}
    \end{subfigure}
    \begin{subfigure}[t]{0.33\textwidth}
    \includegraphics[width=\textwidth]{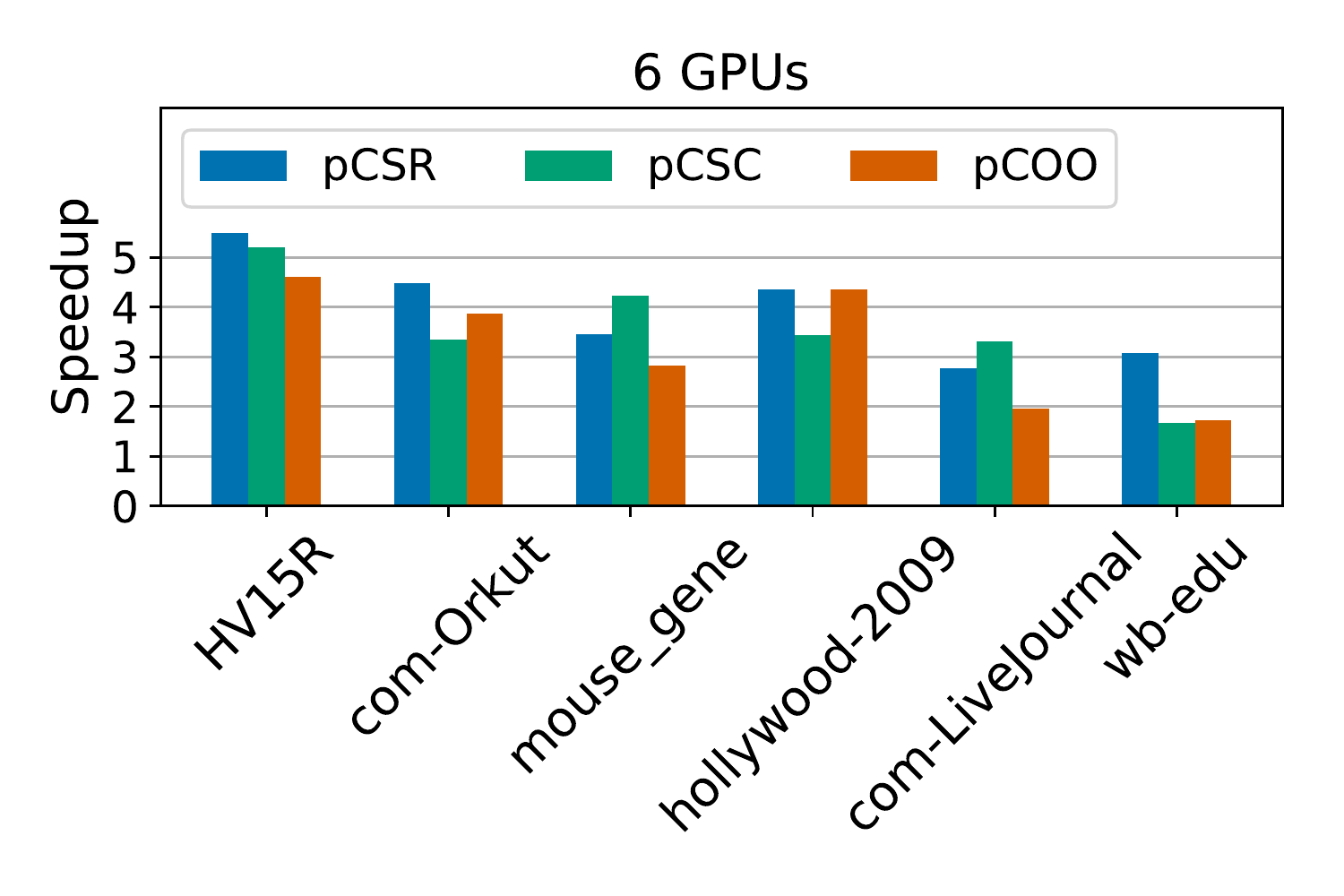}
    \end{subfigure}
    \vspace{-2em}
    \caption{Speedup comparison using different matrices on ORNL Summit.}
     \vspace{-1em}
    \label{multi-smt}
\end{figure*}

\subsection{Partial Results Merging Overhead}
\textbf{Fig. \ref{merg-time}} reports the  time taken to merge the partial results from each GPU. 
Since the baseline design partitions the matrix into row/column block, the cost for merging results based on CSR and COO involves  copying non-overlapped segments of the result vector only. Hence, the overhead is relative low. For CSC representation, each partial result is a vector with the same dimensions as the final result vector, thus need to be added together. For this reason, in this case, the execution time increases linearly with the number of partitions (GPUs).

On the other hand, both pCSR and pCOO  require handling overlapping rows, which results in 
higher cost for merging partial results.
pCSC has the same overhead as the baseline design while merging partial results.

By using GPUs to accelerate the merge step, we are able to reduce the overhead to less than 3.8\% for CSR, 9\% for CSC, and 17\% for COO.
For some cases, optimized merging process brings higher overhead since total execution time is reduced.

\subsection{Effect of NUMA Awareness}
\textbf{Fig. \ref{numa-speedup}} shows the overall speedup of SpMV with and without applying the NUMA-aware design. 
All other optimizations are applied in both cases. 
The cost of copying data in between NUMA nodes are omitted in the results. 
From the figure we can see that, on Summit, the positive impact of NUMA awareness on the performance is clear. The design without NUMA awareness cannot scale well beyond 3 GPUs. However, on DGX-1 we do not consistently observe any NUMA effect.

\subsection{Overall Performance}
We report overall speedup of our SpMV kernel on Summit and DGX-1 machine in \textbf{Fig. \ref{total-speedup}}. From the figure, we observe that the baseline design shows worst performance since it does not perform any optimization to balance the workload. By using our \framework framework and not applying any optimizations, performance improves but still lacks scalability with larger numbers of GPUs. Finally, applying all optimizations discussed earlier, we  achieve near linear speedup with increasing number of GPUs. 
\textbf{Fig. \ref{multi-smt}-\ref{multi-dgx1}} show the speedups on different matrices after applying all optimization techniques.


\section{Discussion}
\label{discussion}

In this section we further discuss our design as well as the potential impact of this work.

\vspace{4pt}\noindent \textbf{Comparing with single GPU works:} Our work lays in between single GPU works and distributed GPU works. Single GPU works that focus on sparse data need to consider both the efficiency of loading data to registers and fine grain parallelism in the thread level. So, it is common to develop a new storage format to facilitate the performance optimization. Our multi-GPU work aims to leverage existing single GPU works and make them scalable beyond one GPU. Thus, one major focus on this work is the compatibility with existing works and sparse data format is the main dominate factor for compatibility. Instead of proposing new formats, we aim to make existing sparse formats scalable across GPUs. Special attention must be made to data partitioning, result merging. Also, we need to explicitly handle memory copies in between CPUs and GPUs considering their inter-connect topology. 

\vspace{4pt}\noindent \textbf{Impact on distributed GPU systems:} Distributed GPU systems are similar to dense multi-GPU systems. So our framework can also support distributed GPU systems. However, since inter-connect bandwidth in between GPU computing nodes is usually limited, special care must be taken when choosing the workload type and data format. For example, SpMV with CSR or COO input brings less communication cost so it is more likely to give relative good scalability on distributed GPU systems.


\vspace{4pt}\noindent \textbf{Benefits to applications:} Enabling scalable sparse operations on multi-GPU systems can potentially benefit many applications that rely on them such as compressed Convolution Neural Network (CNN) \cite{han2015deep}, Graph Convolutional Network \cite{defferrard2016convolutional,hamilton2017inductive}, Graph-based algorithms~\cite{buluc2017graphblas, yang2019graphblast} and many other exascale scientific applications \cite{cook2017proxy}.

\section{Related Works}
\label{related}



In this section, we summarize existing works regarding multi-GPU SpMV, single-GPU SpMV and graph algorithms that adopts can benefit from our sparse data format for multi-GPU execution. 

\vspace{4pt}\noindent\textbf{SpMV on Multi-GPUs:} Yang et al.~\cite{Yang2011} implemented SpMV on a multi-GPU cluster with each CPU node attaching to a single GPU. Each node keeps a local partition of the matrix, while at the end all nodes broadcast their local results to the other nodes. This communication cost of broadcasting is the key factor limiting the scalability.
Our work is distinguished from this work in three aspects: 1) Only partial results are merged with lower overhead; 2) Advanced workload distribution strategy is proposed to handle workload imbalance; 3) Rather than assuming a homogeneous node configuration (using MPI), we optimize our design by addressing NUMA effect. We also want to mention that our intra-node scale-up design is independent of their scale-out design, and thus can be integrated to enable the processing of extremely large matrices.

Kreutzer et al.~\cite{kreutzer2012sparse} proposed an SpMV design based a new sparse matrix storage format called pJDS.
Again, this work also targeted on distributed GPU clusters.
Schubert et al.~\cite{abdelfattah2015high} proposed a multi-GPU SpMV design -- KSPARSE SpMV, based on the Blocked-Sparse-Row format. Their implementation assigned independent tasks across GPUs, making it difficult to conserve workload balancing.
Guo et al.~\cite{guo2016performance} implemented SpMV on multi-GPU systems using CPU multi-threading and concurrent GPU streams, which cannot leverage the high-speed GPU interconnect such as NVLink.

\vspace{2pt}\noindent\textbf{SpMV on Single GPU:} Plenty of research has been done to improve SpMV performance on a single GPU.
Some of them proposed new data formats for facilitating the processing of SpMV, such as CSR5~\cite{liu2015csr5}, BCCOO~\cite{yan2014yaspmv}, SELL-C~\cite{anzt2014implementing}, HYB~\cite{bell2009implementing}, while others applying auto-tuning methodology for improving SpMV performance~\cite{BASMAT,choi2010model,hou2017auto,tan2018design,monakov2010automatically}. Many works also proposed architecture-specialized designs leveraging the unique opportunities from the GPU hardware~\cite{bell2009implementing,hong2018efficient,steinberger2017globally,merrill2016merge,greathouse2014efficient,ashari2014fast}. 
Our work can utilize these fast implementation for single GPU and leverage them in  \framework.

\vspace{2pt}\noindent\textbf{Graph Algorithms:} Graph algorithms are an important class of applications that can directly benefit from the proposed pCSR and pCSC data structures for multi-GPU execution. For example, existing multi-GPU graph analytics frameworks such as Gunrock~\cite{pan2017multi} and Groute~\cite{ben2017groute} essentially partition the graph data in CSR format across multiple GPUs. In addition, GraphBLAS~\cite{buluc2017graphblas} specification expresses graph algorithms as sparse matrix and vector operations on an extended semi-ring algebra, which can be accelerated by GPUs \cite{yang2019graphblast}. Our sparse matrix representation as well as the SpMV design can embark new optimization opportunities for GraphBLAS.


\section{Conclusion}
\label{conclusion}

Sparse linear algebra kernels play a critical role in numerous applications. 
As the volume of data growth exponentially, it will soon become impractical for conducting sparse linear algebra operations in a single GPU due to limited memory capacity and computation performance. In this work, we propose a novel sparse matrix representation framework for multi-GPU systems -- \framework, which aims to help scale sparse linear algebra operations on multiple GPUs while leveraging existing works on a single GPU.
We use SpMV to showcase the efficiency of our framework. 
Evaluation results on an NVIDIA V100-DGX-1 system and the Summit supercomputer show that our SpMV design based on \framework can achieve linear speedup: 5.5$\times$ using 6 GPUs on Summit, and 6.2$\times$ using 8 GPUs on the NVIDIA V100-DGX-1 system. 
\section*{Acknowledgement}
This research was supported by the S-BLAS project under PNNL's High Performance Data Analytics (HPDA) program. This research was supported by the U.S. DOE Office of Science, Office of Advanced Scientific Computing Research, under award 66150: "CENATE - Center for Advanced Architecture Evaluation". The Pacific Northwest National Laboratory is operated by Battelle for the U.S. Department of Energy under Contract DE-AC05-76RL01830.

\bibliographystyle{ACM-Reference-Format}
\bibliography{ref}

\end{document}